\documentclass[a4paper,fleqn,usenatbib]{mnras}
\usepackage{newtxtext,newtxmath}
\usepackage[T1]{fontenc}
\usepackage{ae,aecompl}
\usepackage{graphicx}	
\usepackage{amsmath}	

\usepackage{amssymb}	
\usepackage{mathrsfs}
\usepackage{longtable}
\usepackage{hyperref}
\usepackage[font=small]{caption}
\usepackage{bm}
\usepackage{caption}

\title[A UV census of the environments of SESNe]{A UV census of the environments of stripped-envelope supernovae}

\author[N.-C. Sun et al.]{Ning-Chen Sun$^{1, 2, 3}$\thanks{E-mail: sunnc@ucas.ac.cn}, Justyn R. Maund$^3$ and Paul A. Crowther$^3$\\
1 School of Astronomy and Space Science, University of Chinese Academy of Sciences, 19A Yuquan Road, Shijingshan District, Beijing 100049, China \\
2 National Astronomical Observatories, Chinese Academy of Sciences, 20A Datun Road, Chaoyang District, Beijing 100101, China \\
3 Department of Physics and Astronomy, University of Sheffield, Hicks Building, Hounsfield Road, Sheffield S3 7RH, UK}

\date{Accepted XXX. Received YYY; in original form ZZZ}

\pubyear{2022}

\begin{document}
\label{firstpage}
\pagerange{\pageref{firstpage}--\pageref{lastpage}}
\maketitle

\begin{abstract}

This paper reports an environmental analysis of 41 uniformly-selected stripped-envelope supernovae (SESNe) based on deep ultraviolet-optical images acquired by the Hubble Space Telescope. Young stellar populations are detected in most SN environments and their ages are derived with a hierarchical Bayesian approach. The age distributions are indistinguishable between Type~IIb and Type~Ib while that for Type~Ic is systematically younger. This suggests that the Type~Ic SN progenitors are more massive while the Type~IIb and Type~Ib SNe have very similar progenitor masses. Our result supports a hybrid envelope-stripping mechanism, in which the hydrogen envelopes of the SESN progenitors are stripped via a mass-insensitive process (e.g. binary interaction) while the helium envelopes are stripped via a mass-sensitive process (e.g. stellar wind of the post-binary interaction progenitor). We also provide progenitor constraints for three Type~Ibn SNe and two broad-lined Type~Ic SNe. All these results demonstrate the importance of the very diverse mass-loss processes in the origins of SESNe.

\end{abstract}


\section{Introduction}
\label{intro.sec}

Core-collapse supernovae (SNe) are the spectacular explosions of massive ($>$8~$M_\odot$) stars at the end of their lives. They disperse heavy elements from stars into the space, drive powerful feedback to the interstellar medium, and are associated with the formation of neutron stars or black holes \citep{Heger2003}. They are also important phenomena in multi-messenger astronomy as possible sources of neutrinos, cosmic rays, and gravitational waves \citep{Richardson2022}.

The observational appearances of SNe are significantly affected by the various mass-loss processes of their progenitors. Single stars of $M_{\rm ini}$~= 8--16~$M_\odot$ have relatively weak mass loss and will finally explode as the hydrogen-rich Type~II-P SNe \citep{Smartt2009}. If a star's outer envelope is removed by a strong wind [for Wolf-Rayet (WR) stars with $M_{\rm ini} >$ 25--30~$M_\odot$; \citealt{Crowther2007}] and/or binary interaction \citep{Pods1992}, the supernova explosion will be of Type~IIb (hydrogen lines visible only at early times), Ib (no hydrogen) or Ic (no hydrogen/helium). Some of the Type~Ic SNe, referred to as the ``broad-lined" Type~Ic (Ic-BL\footnote{In the following we shall use ``Type~Ic" to refer to the normal ones, not including Type~Ic-BL unless otherwise specified.}) SNe or ``hypernovae", show very broad lines in their spectra owing to their extremely high explosion energies \citep{Taddia2019}. For some SNe, dense circumstellar material (CSM) can be formed if their progenitors undergo eruptive mass loss shortly before explosion; the later ejecta-CSM interaction can produce strong and narrow emission lines, and such a SN will be of Type~IIn, Ibn or Icn depending on the chemical composition of the progenitor and its CSM \citep{Smith2014, Smith2017}.

Arising from stripped progenitors, the hydrogen-poor SNe are also referred to as the ``stripped-envelope" (SE) SNe, including the above-mentioned Types~IIb, Ib, Ic, Ic-BL, Ibn and Icn. Compared with the Type~II-P SNe, the search for SESN progenitors on pre-explosion images has proven to be much more difficult. Currently, the progenitors have only been detected for 5 Type~IIb SNe (1993J, 2008ax, 2011dh, 2013df, 2016gkg; \citealt{Aldering1994, Crockett2008, Maund2011, sn2013df.ref, Kilpatrick2017, Tartaglia2017}), 2 Type~Ib SNe (iPTF13bvn and 2019yvr; \citealt{iptf13bvn.ref, Eldridge2015, Kilpatrick2021, Sun2022a}) and 1 Type~Ic SN (2017ein; \citealt{VanDyk2018, Xiang2019}). Binary companions have (possibly) been detected for the Type~IIb SN~1993J \citep{Maund2004}, SN~2001ig \citep{Ryder2018} and SN~2011dh \citep{Maund2019}, the Type~Ib SN~2013ge \citep{Fox2022} and SN~2019yvr \citep{Sun2022a}, and the Type~Ibn SN~2006jc \citep{Maund2016b, Sun2020a}. In addition, host star clusters (which can be used to infer the progenitor properties) have been found for the Type~Ib SN~2014C \citep{sn2014c.ref2, Sun2020b} and the Type~Ic SN~2020oi \citep{Gagliano2022}.

The limited number of these direct detections poses a major challenge to our understanding of the progenitors of SESNe and their pre-SN mass loss. This motivates astronomers to explore alternative methods to constrain the progenitors for SESNe, such as measuring the line strengths of heavy elements in the nebular-phase spectra \citep[e.g.][]{Fang2019, Fang2022} and deriving the ejecta masses by fitting light curves \citep[e.g.][]{Lyman2016, Taddia2019}. Environmental analysis is another powerful method to study SN progenitors, since most massive stars form in groups and stars in each group share very similar ages and metallicities \citep[e.g.][]{Anderson2012, Kangas2013, Galbany2018, K2018, Schady2019, Xiao2019}. Generally speaking, SNe from higher-mass progenitors will more likely to be associated with tracers of recent star formation (e.g. H$\alpha$ emission, dust, young stars, etc.), while those from lower-mass progenitors may reside in much older environments.

Among the various techniques of environmental analysis, deep and high-spatial resolution imaging conducted by the Hubble Space Telescope (HST) can directly probe the stellar populations in the SN environments, and the stellar ages can be derived by fitting stellar isochrones on the colour-magnitude diagrams (CMDs) \citep[e.g.][]{Maund2016, Maund2017, Maund2018, Williams2019, Sun2021, Sun2022a}. Compared with other techniques, this approach provides (relatively) more accurate age measurements. The past works have, however, largely relied on heterogeneous collections of archival observations and could suffer from statistical biases that are difficult to assess. Moreover, the age-dependance of the colours of the very young and massive stars, often seen in SESN environments, is most sensitive at ultraviolet (UV) wavelengths. For many SESNe, however, observations have been performed only at optical wavelengths.

In this paper, we report a homogeneous set of deep UV-optical imaging observations of a uniformly-selected sample of 41 SESNe in the Local Universe. Taking advantage of the UV filter and the deep detection limits, we try to probe the very young stellar populations in their environments and acquire an accurate determination of their ages. We also make statistical comparisons between the different SN types. Our aim is to reveal their possible progenitor channels and the roles of the various mass-loss processes in their origins.

This paper is structured as follows: in Section~\ref{sample.sec} we describe the sample selection, observation, photometry and analysis; Section~\ref{results.sec} provides the results and discussions for each individual SN types; and we finally close the paper with a summary and our conclusions in Section~\ref{summary.sec}.

\section{Sample and analysis}
\label{sample.sec}

\begin{table*}
\center
\caption{Sample of SESNe analyzed in this work}
\begin{tabular}{ccccccl}
\hline
\hline
Target & Type & Host galaxy & Distance & Inclination & Galactic extinction & References \\
 & & & $D$ (Mpc) & $\theta$ (deg) & $A_V^{\rm MW}$ (mag) & \\
\hline
1996cb & IIb & NGC~3510 & 16.8 (2.3) & 78.1 & 0.08 & \citealt{sn1996cb.ref} \\
2003bg & IIb & ESO~420-g009 & 17.7 (1.2) & 41.7 & 0.06 & \citealt{sn2003bg.ref1, sn2003bg.ref2} \\
2004C & IIb & NGC~3683 & 32.1 (4.4) & 69.0 & 0.04 & \citealt{sn2004c.ref} \\
2005ae & IIb & ESO~209-g9 & 13.6 (1.9) & 90.0 & 0.69 & \citealt{sn2005ae.ref} \\
2008bo & IIb & NGC~6643 & 20.9 (2.9) & 62.7 & 0.16 & \citealt{sn2008bo.ref} \\
2009dq & IIb & IC~2554 & 16.4 (1.1) & 70.8 & 0.55 & \citealt{sn2009dq.ref} \\
2009gj & IIb & NGC~134 & 19.2 (2.7) & 90.0 & 0.05 & \citealt{sn2009gj.ref} \\
2009mk & IIb & ESO~293-g34 & 20.7 (1.4) & 74.6 & 0.04 & \citealt{sn2009mk.ref} \\
2010as & IIb & NGC~6000 & 31.0 (2.1) & 30.7 & 0.46 & \citealt{sn2010as.ref} \\
2011hs & IIb & IC~5267 & 20.9 (4.8) & 47.8 & 0.03 & \citealt{sn2011hs.ref} \\
2013df & IIb & NGC~4414 & 17.9 (1.4) & 56.6 & 0.05 & \citealt{sn2013df.ref} \\
2015Y & IIb & NGC~2735 & 54.7 (11.3) & 74.2 & 0.10 & \citealt{sn2015y.ref} \\
2016bas & IIb & ESO~163-g11 & 50.6 (9.3) & 70.9 & 0.40 & \citealt{sn2016bas.ref} \\
\hline
1996N & Ib & NGC~1398 & 28.6 (5.9) & 47.5 & 0.04 & \citealt{sn1996n.ref} \\
2000ds & Ib & NGC~2768 & 22.2 (2.9) & 90.0 & 0.12 & \citealt{sn2000ds.ref} \\
2001B & Ib & IC~391 & 25.2 (1.7) & 18.1 & 0.34 & \citealt{sn2001b.ref} \\
2004ao & Ib & UGC~10862 & 29.6 (2.0) & 53.4 & 0.27 & \citealt{sn2004ao.ref} \\
iPTF13bvn & Ib & NGC~5806 & 27.0 (3.7) & 60.4 & 0.14 & \citealt{iptf13bvn.ref} \\
2014C & Ib & NGC~7331 & 14.1 (1.0) & 70.0 & 0.24 & \citealt{sn2014c.ref1, sn2014c.ref2} \\
2014df & Ib & NGC~1448 & 17.2 (1.3) & 86.4 & 0.04 & \citealt{sn2014df.ref} \\
2015Q & Ib & NGC~3888 & 39.3 (2.7) & 36.3 & 0.03 & \citealt{sn2015q.ref} \\
2016cdd & Ib & ESO~218-g8 & 28.7 (2.0) & 90.0 & 0.49 & \citealt{sn2016cdd.ref} \\
\hline
1990W & Ic & NGC~6221 & 11.9 (1.6) & 50.9 & 0.44 & \citealt{sn1990w.ref1, sn1990w.ref2} \\
2000ew & Ic & NGC~3810 & 15.9 (2.2) & 48.2 & 0.12 & \citealt{sn2000ew.ref} \\
2001ci & Ic & NGC~3079 & 20.6 (3.8) & 90.0 & 0.03 & \citealt{sn2001ci.ref} \\
2004bm & Ic & NGC~3437 & 24.2 (3.3) & 72.8 & 0.05 & \citealt{sn2004bm.ref} \\
2004gk & Ic & IC~3311 & 20.6 (2.8) & 90.0 & 0.08 & \citealt{sn2004gk.ref} \\
2005at & Ic & NGC~6744 & 9.0 (0.7) & 53.5 & 0.11 & \citealt{sn2005at.ref1, sn2005at.ref2} \\
2005aw & Ic & IC~4837a & 41.6 (2.9) & 90.0 & 0.16 & \citealt{sn2005aw.ref} \\
2009em & Ic & NGC~157 & 12.1 (2.2) & 61.8 & 0.12 & \citealt{sn2009em.ref1, sn2009em.ref2} \\
2012cw & Ic & NGC~3166 & 22.0 (1.5) & 56.2 & 0.08 & \citealt{sn2012cw.ref} \\
2012fh & Ic & NGC~3344 & 9.8 (0.9) & 18.7 & 0.09 & \citealt{sn2012fh.ref} \\
\hline
1991N & Ib/Ic & NGC~3310 & 19.2 (1.3) & 16.1 & 0.06 & \citealt{sn1991n.ref} \\
1995F & Ib/Ic & NGC~2726 & 55.7 (10.3) & 90.0 & 0.10 & \citealt{sn1995f.ref} \\
2005V & Ib/Ic & NGC~2146 & 16.7 (1.2) & 37.4 & 0.26 & \citealt{sn2005v.ref1, sn2005v.ref2} \\
2013ge & Ib/Ic & NGC~3287 & 14.6 (2.0) & 75.3 & 0.06 & \citealt{sn2013ge.ref} \\
\hline
2006jc & Ibn & UGC~4904 & 27.8 (1.9) & 52.2 & 0.05 & \citealt{sn2006jc.ref1, sn2006jc.ref2} \\
2015G & Ibn & NGC~6951 & 18.8 (1.7) & 50.8 & 0.99 & \citealt{sn2015g.ref} \\
2015U & Ibn & NGC~2388 & 60.3 (11.1) & 52.6 & 0.15 & \citealt{sn2015u.ref} \\
\hline
1997dq & Ic-BL & NGC~3810 & 15.9 (2.2) & 48.2 & 0.12 & \citealt{sn1997dq.ref} \\
2002ap & Ic-BL & NGC~628 & 9.8 (0.9) & 19.8 & 0.19 & \citealt{sn2002ap.ref} \\
\hline
\end{tabular}
\label{sample.tab}
\end{table*}

\begin{table*}
\center
\caption{Parameters derived from the stellar population fitting for the SN environments. $N_{\rm m}$: number of age components to fit the observed stars; $A_V^{\rm int}$ and d$A_V^{\rm int}$: the mean value and standard deviation of the internal extinctions from the host galaxy; log($t_i$/yr) and $w_i$ ($i$ = 1, 2, 3): mean stellar log-age and weight for the $i$-th age component; log($t_{\rm lim}$/yr): age limit older than which a stellar population will fall below the detection limits. The last column gives the numbers of data points used in the fitting, with $N_{\rm both}$, $N_{\rm F300X}$ and $N_{\rm F475X}$ corresponding to those detected in both bands, only in the F300X band and only in the F475X band, respectively. The errors are propagated from photometric uncertainties.}
\begin{tabular}{ccccccccccccc}
\hline
\hline
Target & Type & $N_{\rm m}$ & $A_V^{\rm int}$ & ${\rm d}A_V^{\rm int}$ & log($t_1$/yr) & $w_1$ & log($t_2$/yr) & $w_2$ & log($t_3$/yr) & $w_3$ & log($t_{\rm lim}$/yr) & $N_{\rm both}$, $N_{\rm F300X}$,  \\
& & & (mag) & (mag) & & & & & & & & $N_{\rm F475X}$ \\
\hline
1996cb & IIb & 2 & 0.04$_{-0.01}^{+0.02}$ & 0.01$_{-0.00}^{+0.01}$ & 6.86$_{-0.03}^{+0.03}$ & 0.64 & 7.21$_{-0.03}^{+0.03}$ & 0.36 & ...  & ...  & 7.5
& 25, 23, 16
\\
2003bg & IIb & 2 & 0.05$_{-0.02}^{+0.03}$ & 0.02$_{-0.00}^{+0.01}$ & 6.82$_{-0.08}^{+0.04}$ & 0.72 & 7.31$_{-0.11}^{+0.12}$ & 0.28 & ...  & ...  & 7.5
& 10, 10, 5
\\
2004C & IIb & 1 & 0.77$_{-0.12}^{+0.11}$ & 0.30$_{-0.07}^{+0.06}$ & 6.74$_{-0.02}^{+0.02}$ & 1.00 & ...  & ...  & ...  & ...  & 6.9
& 17, 2, 37
\\
2005ae & IIb  & 0 & ... & ... & ... & ... & ... & ... & ... & ... & 7.3 \\
2008bo & IIb & 1 & 1.39$_{-0.09}^{+0.09}$ & 0.67$_{-0.06}^{+0.06}$ & 6.60$_{-0.02}^{+0.03}$ & 1.00 & ...  & ...  & ...  & ...  & 6.8
& 43, 12, 38
\\
2009dq & IIb & 1 & 1.09$_{-0.05}^{+0.05}$ & 0.33$_{-0.03}^{+0.03}$ & 6.72$_{-0.01}^{+0.01}$ & 1.00 & ...  & ...  & ...  & ...  & 6.9
& 74, 5, 195
\\
2009gj & IIb  & 0 & ... & ... & ... & ... & ... & ... & ... & ... & 7.5 \\
2009mk & IIb & 1 & 0.28$_{-0.06}^{+0.06}$ & 0.05$_{-0.05}^{+0.03}$ & 6.94$_{-0.02}^{+0.02}$ & 1.00 & ...  & ...  & ...  & ...  & 7.3
& 22, 8, 21
\\
2010as & IIb & 2 & 0.17$_{-0.07}^{+0.10}$ & 0.03$_{-0.02}^{+0.03}$ & 6.57$_{-0.03}^{+0.03}$ & 0.38 & 6.76$_{-0.01}^{+0.01}$ & 0.62 & ...  & ...  & 7.0
& 56, 5, 52
\\
2011hs & IIb & 1 & 0.26$_{-0.12}^{+0.14}$ & 0.05$_{-0.05}^{+0.06}$ & 6.90$_{-0.06}^{+0.12}$ & 1.00 & ...  & ...  & ...  & ...  & 7.4
& 3, 1, 4
\\
2013df & IIb & 2 & 0.13$_{-0.04}^{+0.05}$ & 0.03$_{-0.02}^{+0.02}$ & 6.92$_{-0.02}^{+0.02}$ & 0.62 & 7.14$_{-0.04}^{+0.04}$ & 0.38 & ...  & ...  & 7.4
& 19, 24, 19
\\
2015Y & IIb  & 0 & ... & ... & ... & ... & ... & ... & ... & ... & 7.0 \\
2016bas & IIb & 1 & 0.08$_{-0.03}^{+0.04}$ & 0.02$_{-0.01}^{+0.01}$ & 6.66$_{-0.01}^{+0.01}$ & 1.00 & ...  & ...  & ...  & ...  & 6.8
& 25, 1, 13
\\
\hline
1996N & Ib  & 0 & ... & ... & ... & ... & ... & ... & ... & ... & 7.3 \\
2000ds & Ib  & 0 & ... & ... & ... & ... & ... & ... & ... & ... & 7.4 \\
2001B & Ib & 1 & 0.06$_{-0.02}^{+0.03}$ & 0.02$_{-0.00}^{+0.01}$ & 6.83$_{-0.01}^{+0.01}$ & 1.00 & ...  & ...  & ...  & ...  & 7.2
& 28, 8, 23
\\
2004ao & Ib & 1 & 0.75$_{-0.11}^{+0.11}$ & 0.47$_{-0.08}^{+0.08}$ & 6.55$_{-0.04}^{+0.04}$ & 1.00 & ...  & ...  & ...  & ...  & 6.9
& 19, 5, 6
\\
iPTF13bvn & Ib & 1 & 0.78$_{-0.13}^{+0.13}$ & 0.05$_{-0.05}^{+0.06}$ & 6.71$_{-0.04}^{+0.04}$ & 1.00 & ...  & ...  & ...  & ...  & 7.0
& 6, 0, 2
\\
2014C & Ib & 1 & 0.62$_{-0.04}^{+0.04}$ & 0.03$_{-0.03}^{+0.03}$ & 6.93$_{-0.01}^{+0.01}$ & 1.00 & ...  & ...  & ...  & ...  & 7.3
& 5, 3, 40
\\
2014df & Ib & 1 & 0.55$_{-0.06}^{+0.06}$ & 0.25$_{-0.05}^{+0.04}$ & 6.91$_{-0.02}^{+0.02}$ & 1.00 & ...  & ...  & ...  & ...  & 7.3
& 21, 14, 21
\\
2015Q & Ib & 1 & 0.06$_{-0.04}^{+0.07}$ & 0.02$_{-0.01}^{+0.02}$ & 6.74$_{-0.03}^{+0.03}$ & 1.00 & ...  & ...  & ...  & ...  & 7.1
& 4, 2, 1
\\
2016cdd & Ib & 2 & 0.26$_{-0.04}^{+0.04}$ & 0.14$_{-0.03}^{+0.03}$ & 6.63$_{-0.01}^{+0.01}$ & 0.75 & 6.76$_{-0.03}^{+0.02}$ & 0.25 & ...  & ...  & 6.9
& 65, 7, 22
\\
\hline
1990W & Ic & 2 & 0.78$_{-0.04}^{+0.04}$ & 0.34$_{-0.03}^{+0.03}$ & 6.62$_{-0.02}^{+0.01}$ & 0.68 & 6.84$_{-0.02}^{+0.02}$ & 0.32 & ...  & ...  & 7.2
& 176, 32, 60
\\
2000ew & Ic & 2 & 0.78$_{-0.07}^{+0.06}$ & 0.28$_{-0.05}^{+0.04}$ & 6.61$_{-0.02}^{+0.03}$ & 0.50 & 6.84$_{-0.01}^{+0.01}$ & 0.50 & ...  & ...  & 7.2
& 128, 21, 57
\\
2001ci & Ic  & 0 & ... & ... & ... & ... & ... & ... & ... & ... & 7.4 \\
2004bm & Ic & 2 & 1.20$_{-0.06}^{+0.05}$ & 0.21$_{-0.07}^{+0.05}$ & 6.56$_{-0.03}^{+0.03}$ & 0.25 & 6.75$_{-0.01}^{+0.01}$ & 0.75 & ...  & ...  & 6.9
& 34, 2, 76
\\
2004gk & Ic & 1 & 0.62$_{-0.16}^{+0.12}$ & 0.23$_{-0.12}^{+0.07}$ & 6.82$_{-0.03}^{+0.04}$ & 1.00 & ...  & ...  & ...  & ...  & 7.2
& 16, 8, 13
\\
2005at & Ic & 2 & 0.61$_{-0.07}^{+0.07}$ & 0.34$_{-0.05}^{+0.05}$ & 6.59$_{-0.11}^{+0.10}$ & 0.76 & 7.28$_{-0.07}^{+0.05}$ & 0.24 & ...  & ...  & 7.6
& 57, 24, 13
\\
2005aw & Ic & 1 & 0.58$_{-0.13}^{+0.12}$ & 0.05$_{-0.06}^{+0.06}$ & 6.75$_{-0.02}^{+0.02}$ & 1.00 & ...  & ...  & ...  & ...  & 6.9
& 4, 0, 13
\\
2009em & Ic & 3 & 0.39$_{-0.04}^{+0.04}$ & 0.19$_{-0.02}^{+0.03}$ & 6.72$_{-0.04}^{+0.03}$ & 0.56 & 6.93$_{-0.05}^{+0.04}$ & 0.30 & 7.18$_{-0.04}^{+0.05}$ & 0.14 & 7.5
& 102, 37, 21
\\
2012cw & Ic  & 0 & ... & ... & ... & ... & ... & ... & ... & ... & 7.4 \\
2012fh & Ic & 2 & 0.48$_{-0.03}^{+0.03}$ & 0.24$_{-0.02}^{+0.02}$ & 6.53$_{-0.06}^{+0.05}$ & 0.85 & 7.49$_{-0.03}^{+0.04}$ & 0.15 & ...  & ...  & 7.6
& 126, 39, 30
\\
\hline
1991N & Ib/Ic & 2 & 0.31$_{-0.02}^{+0.03}$ & 0.13$_{-0.01}^{+0.01}$ & 6.58$_{-0.01}^{+0.01}$ & 0.86 & 6.76$_{-0.01}^{+0.01}$ & 0.14 & ...  & ...  & 7.1
& 345, 35, 34
\\
1995F & Ib/Ic  & 0 & ... & ... & ... & ... & ... & ... & ... & ... & 6.9 \\
2005V & Ib/Ic  & 0 & ... & ... & ... & ... & ... & ... & ... & ... & 7.3 \\
2013ge & Ib/Ic & 1 & 1.06$_{-0.26}^{+0.25}$ & 0.76$_{-0.17}^{+0.12}$ & 6.59$_{-0.22}^{+0.20}$ & 1.00 & ...  & ...  & ...  & ...  & 7.2
& 5, 5, 4
\\
\hline
2006jc & Ibn  & 0 & ... & ... & ... & ... & ... & ... & ... & ... & 7.3 \\
2015G & Ibn  & 0 & ... & ... & ... & ... & ... & ... & ... & ... & 7.2 \\
2015U & Ibn  & 0 & ... & ... & ... & ... & ... & ... & ... & ... & 6.9 \\
\hline
1997dq & Ic-BL & 1 & 0.65$_{-0.16}^{+0.17}$ & 0.40$_{-0.13}^{+0.12}$ & 6.94$_{-0.05}^{+0.05}$ & 1.00 & ...  & ...  & ...  & ...  & 7.3
& 6, 7, 9
\\
2002ap & Ic-BL & 1 & 0.23$_{-0.07}^{+0.10}$ & 0.08$_{-0.10}^{+0.06}$ & 6.96$_{-0.05}^{+0.04}$ & 1.00 & ...  & ...  & ...  & ...  & 7.6
& 7, 7, 2
\\
\hline
\end{tabular}
\label{pop.tab}
\end{table*}

\begin{figure*}
\includegraphics[width=0.49\linewidth]{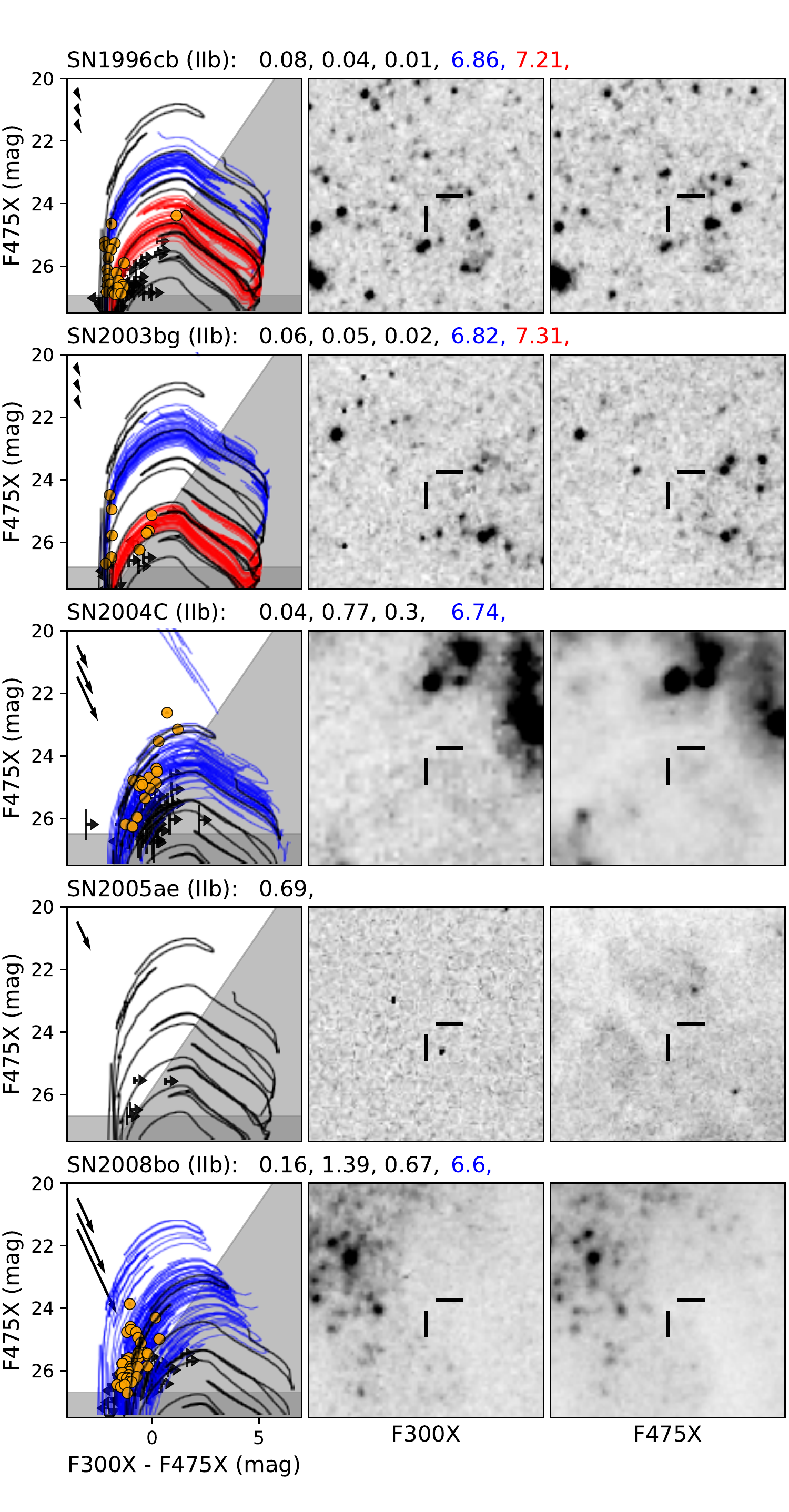}
\includegraphics[width=0.49\linewidth]{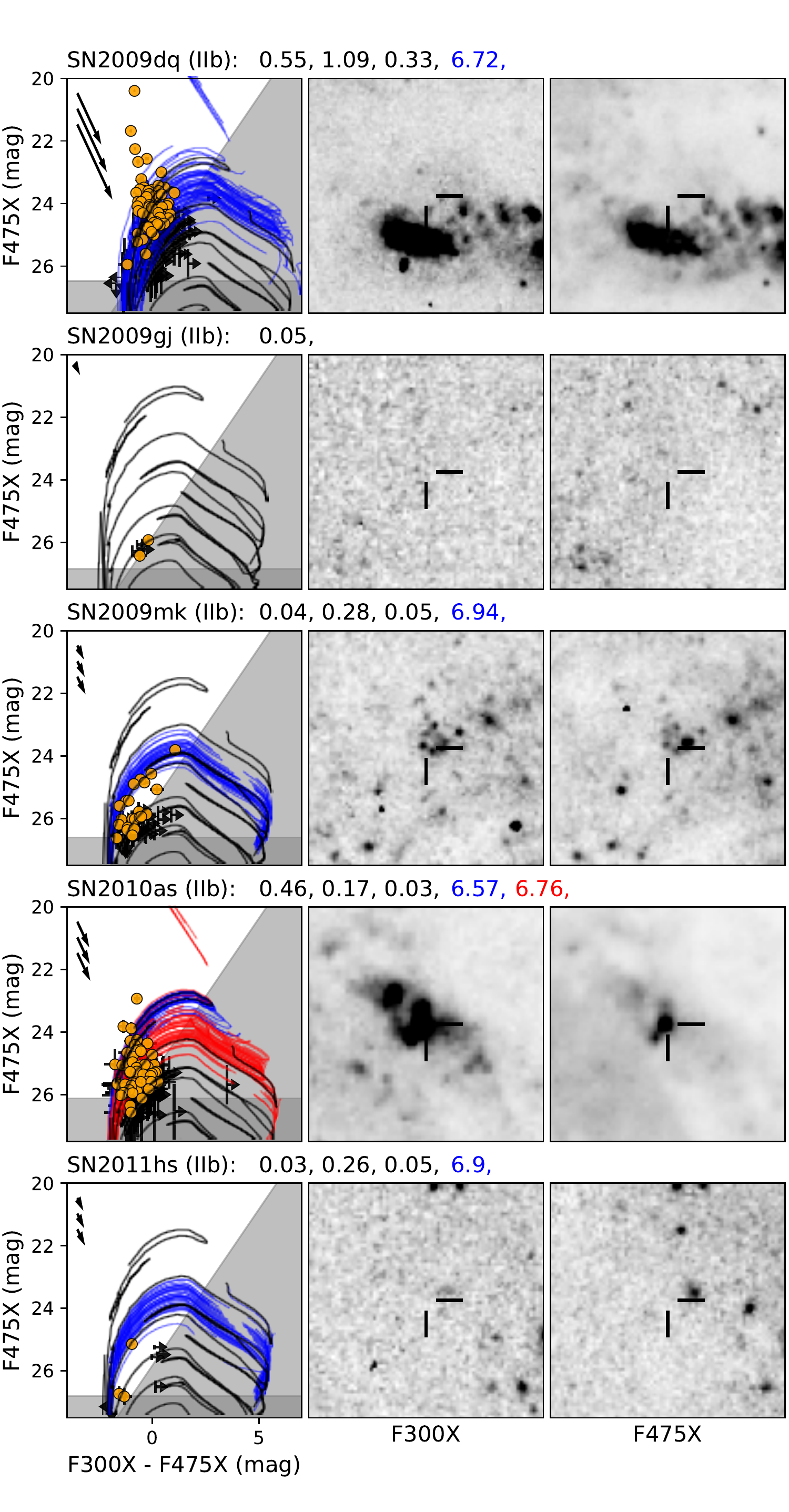}
\caption{CMDs and HST images of the Type~IIb SN environments as listed in Table~\ref{sample.tab}. The HST images have sizes of 300~pc $\times$ 300~pc and are centred on the SN positions (indicated by the crosshairs). The circular region we defined for environmental analysis has a radius of 150~pc, i.e. a diameter equal to the box size. In the CMDs, the grey-shaded regions show the average detection limits for a random position in the SN environments, below which the 5$\sigma$ detection probability falls below 50\%. Overlaid are \textsc{parsec} (v1.2S) stellar isochrones; the colored ones correspond to the fitted model stellar populations, with blue, red and green colours used for populations of increasing ages, and, for each age component, we show 30 random realisations derived from their age and extinction spreads; for reference, we also show black isochrones with log-ages of 6.6, 6.8, 7.0, ... in equal spacing of 0.2~dex, all reddened with the Galactic and mean internal extinction. The title for each SN lists values of the Galactic extinction $A_V^{\rm MW}$, and if a stellar population fitting is performed, the mean internal extinction $A_V^{\rm int}$, internal extinction standard deviation d$A_V^{\rm int}$, and the mean log-ages of the model stellar populations log($t_i$/yr) ($i$ = 1, 2, 3). The three arrows in the upper-left corners show the reddening vectors with lengths corresponding to $A_V^{\rm MW}$ + $A_V^{\rm int}$ ($\pm$ d$A_V^{\rm int}$). }
\label{IIb.fig}
\end{figure*}

\begin{figure*}
\includegraphics[width=0.49\linewidth]{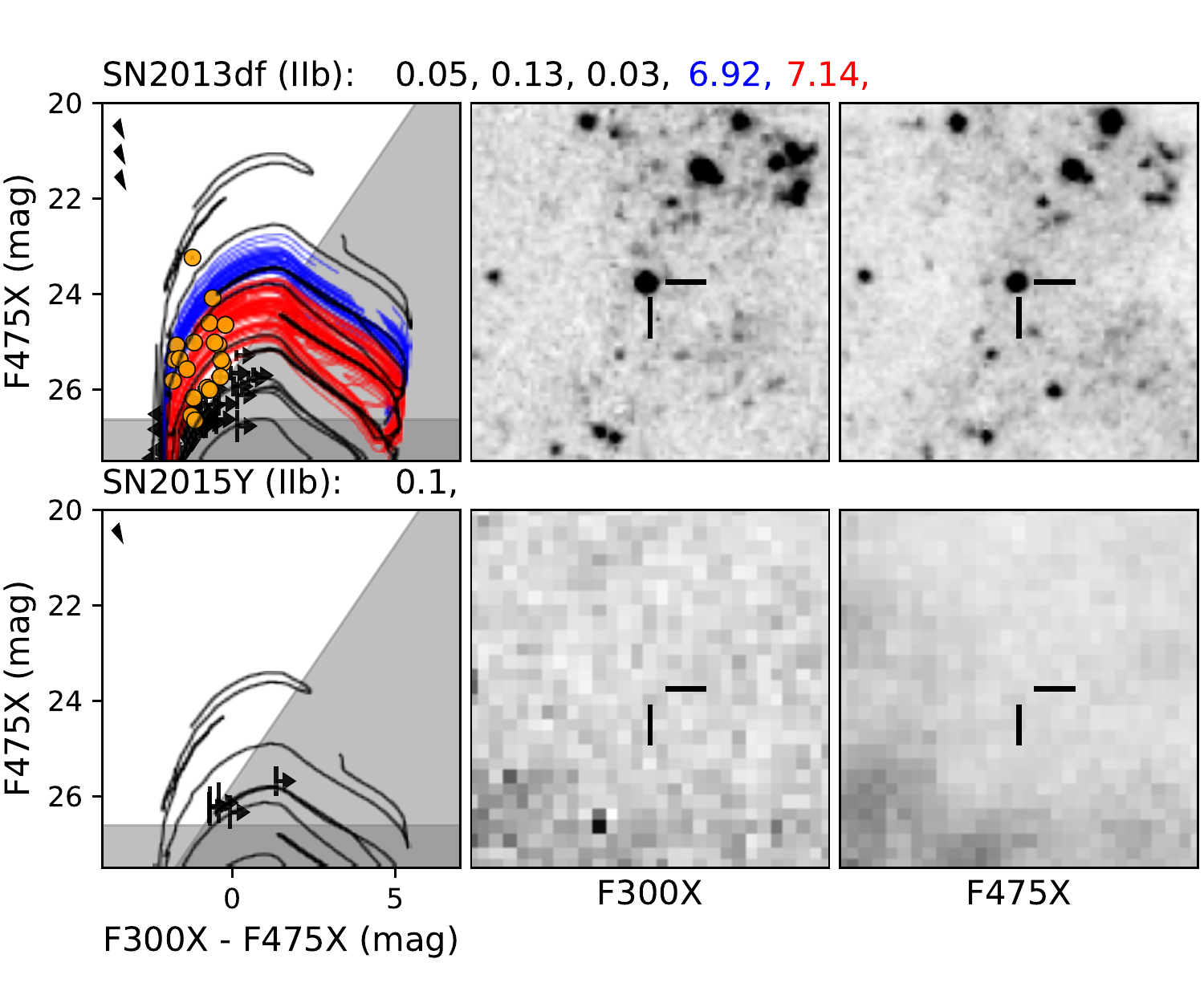}
\includegraphics[width=0.49\linewidth]{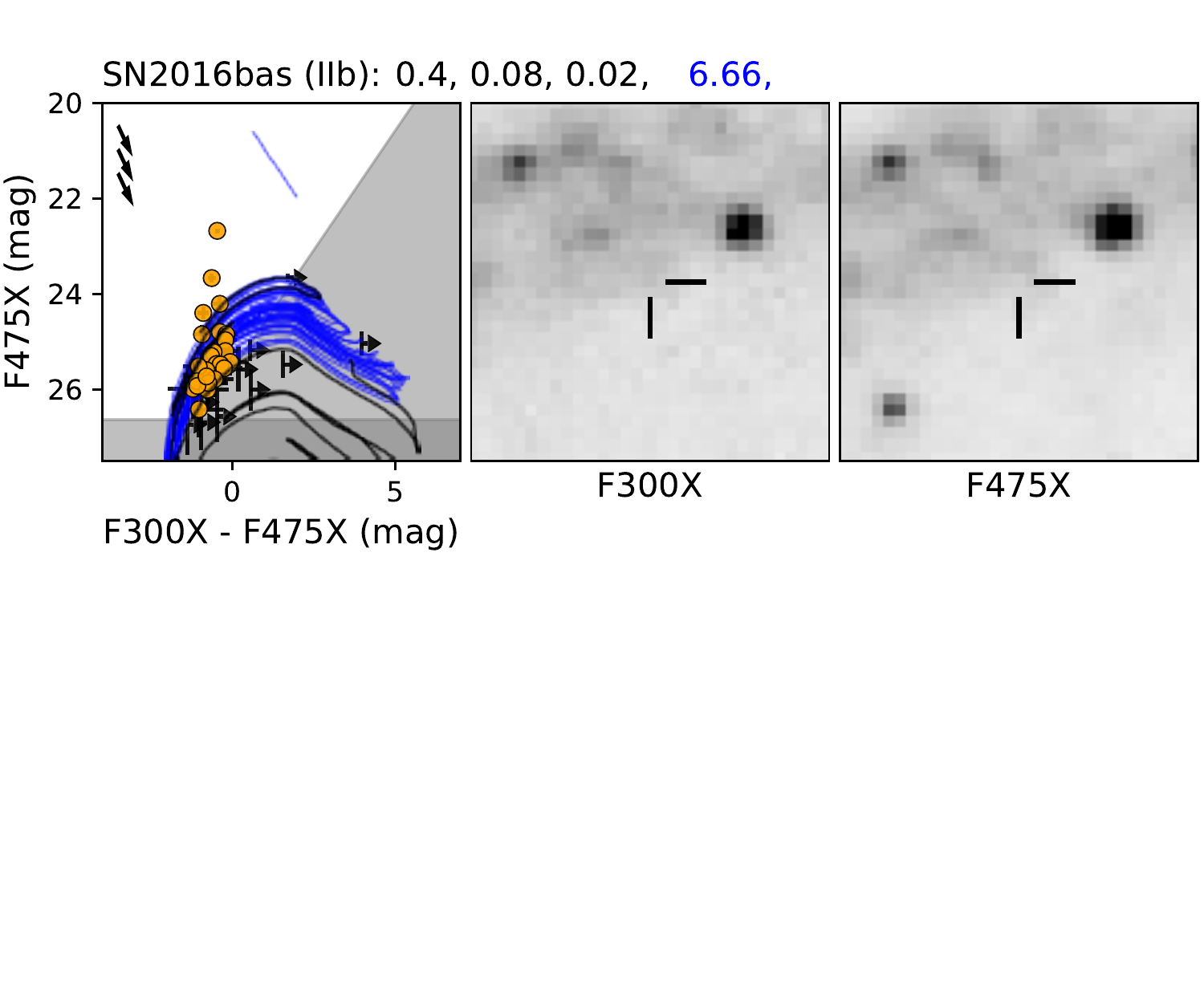}
\smallskip \centerline{\textbf{Figure 1} -- continued.} \smallskip \noindent
\end{figure*}

\begin{figure*}
\includegraphics[width=0.49\linewidth]{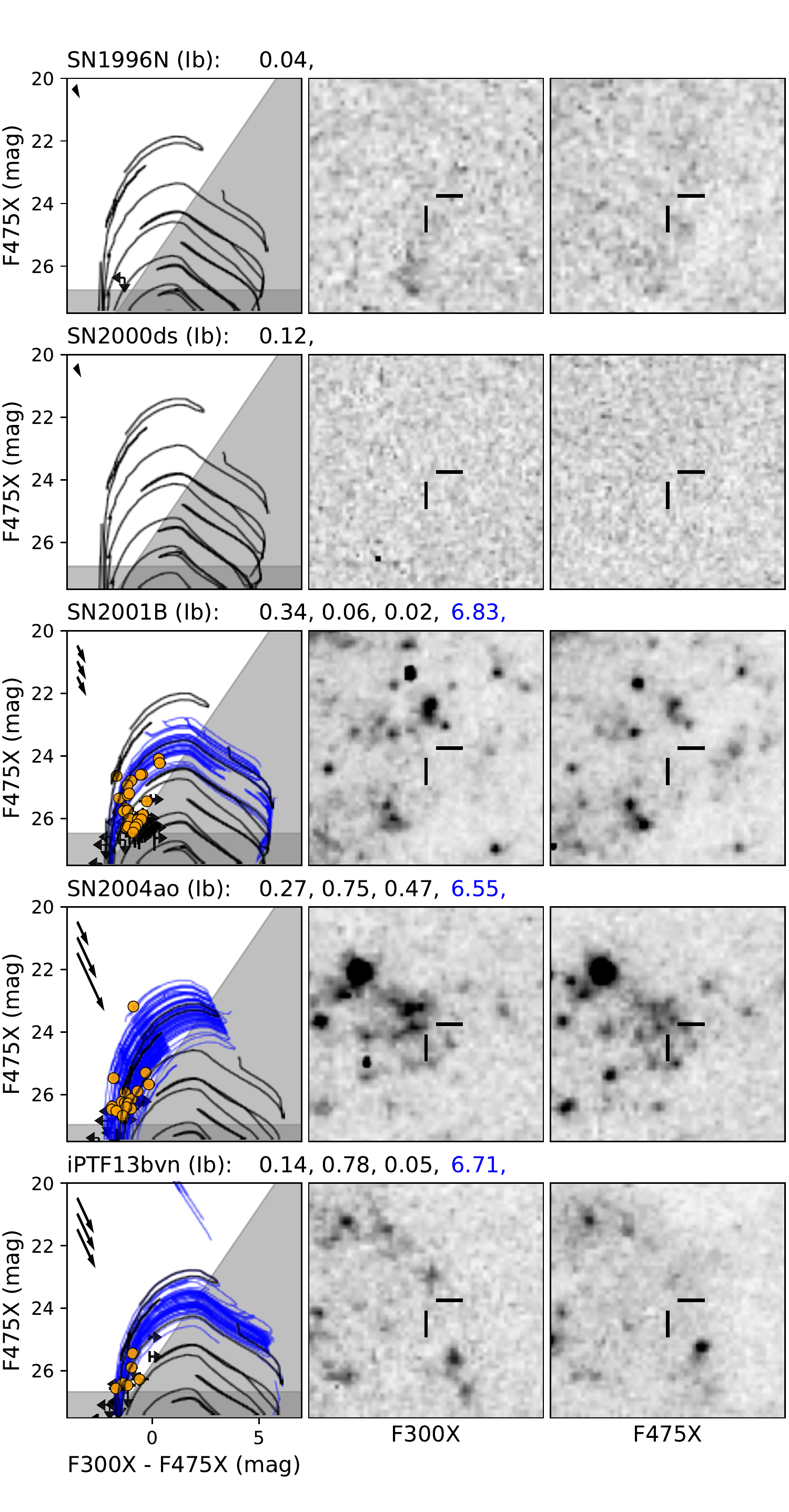}
\includegraphics[width=0.49\linewidth]{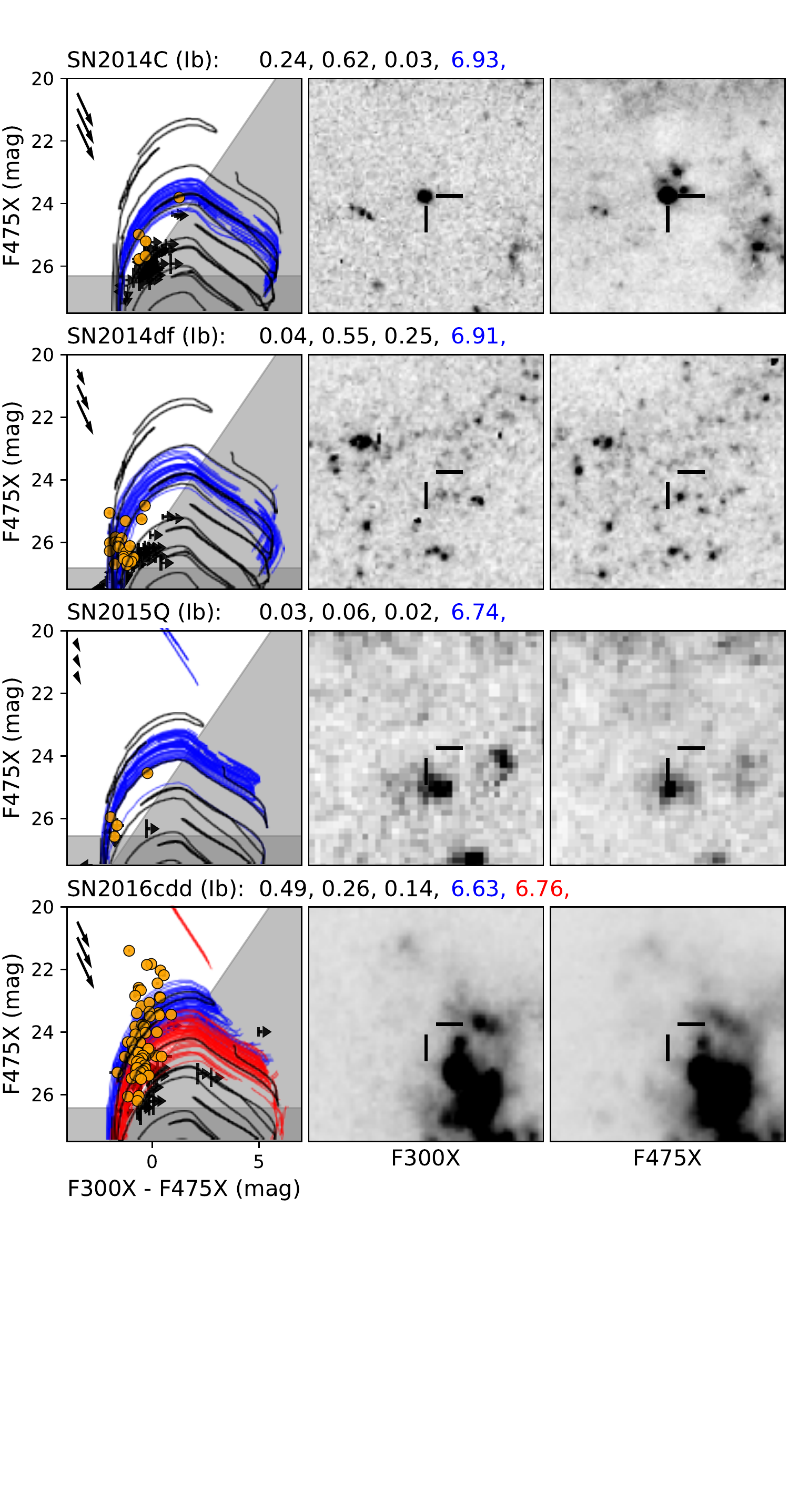}
\caption{Same as Fig.~\ref{IIb.fig} but for Type~Ib SNe.}
\label{Ib.fig}
\end{figure*}

\begin{figure*}
\includegraphics[width=0.49\linewidth]{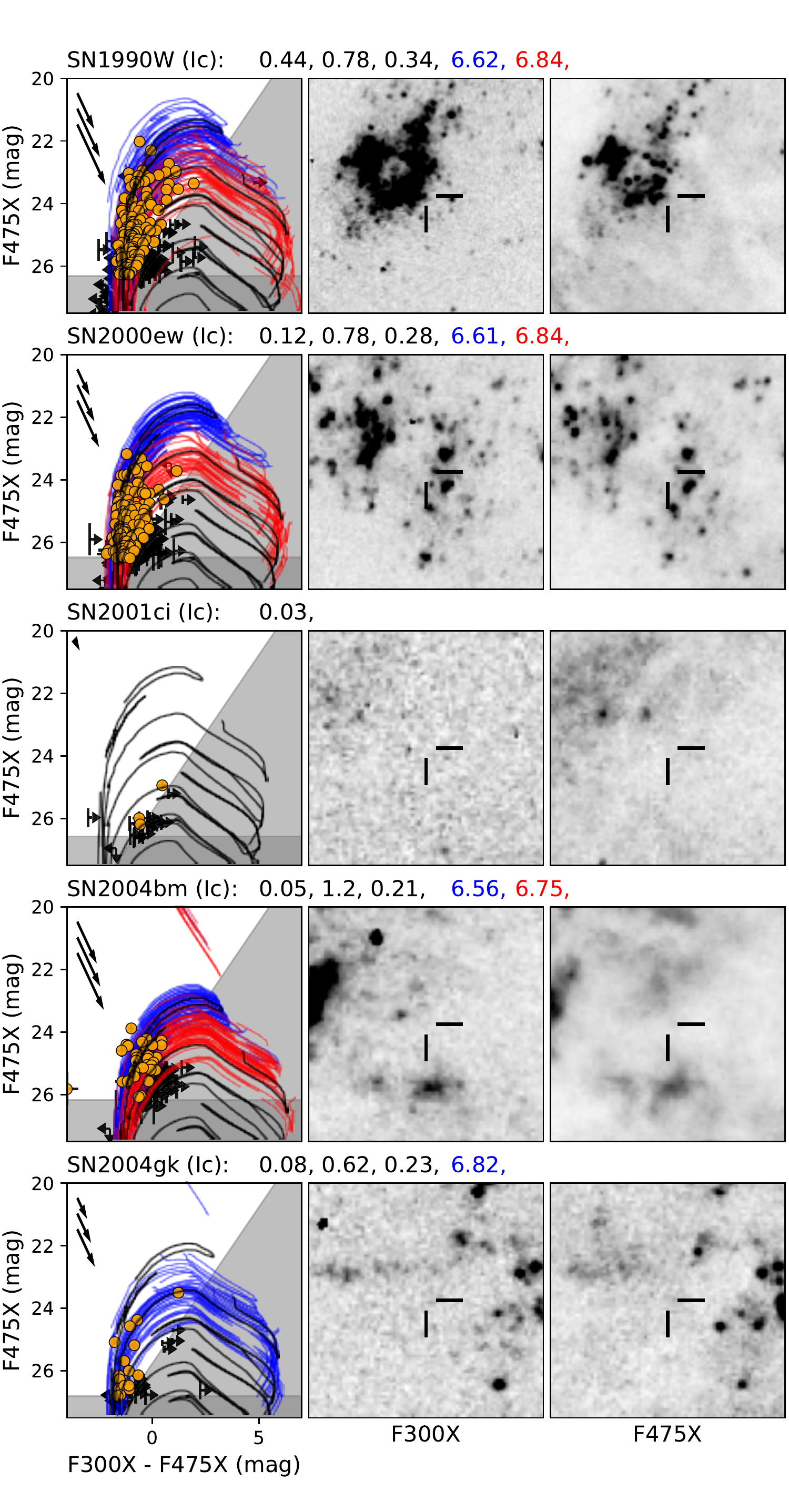}
\includegraphics[width=0.49\linewidth]{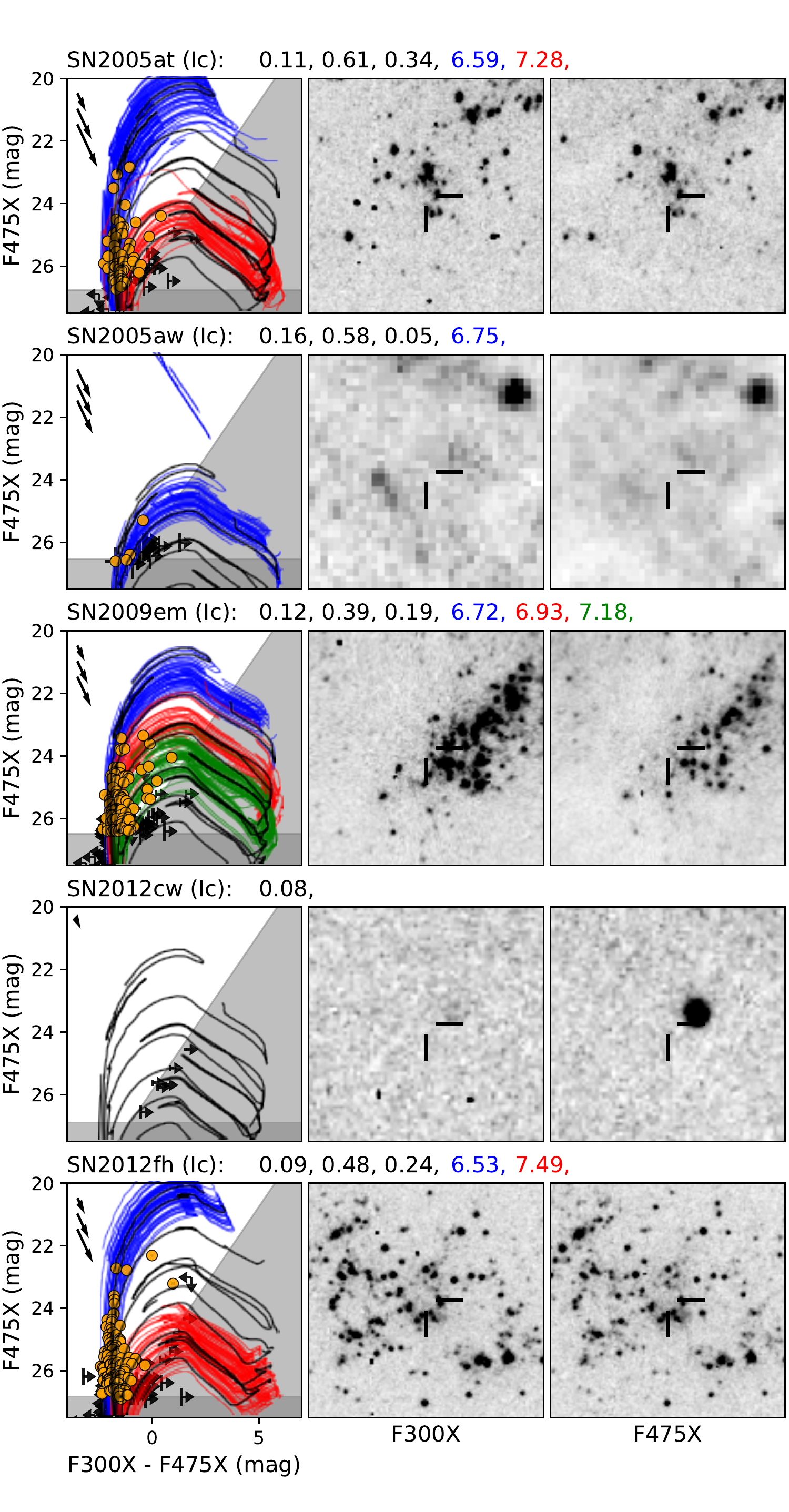}
\caption{Same as Fig.~\ref{IIb.fig} but for Type~Ic SNe.}
\label{Ic.fig}
\end{figure*}

\begin{figure*}
\includegraphics[width=0.49\linewidth]{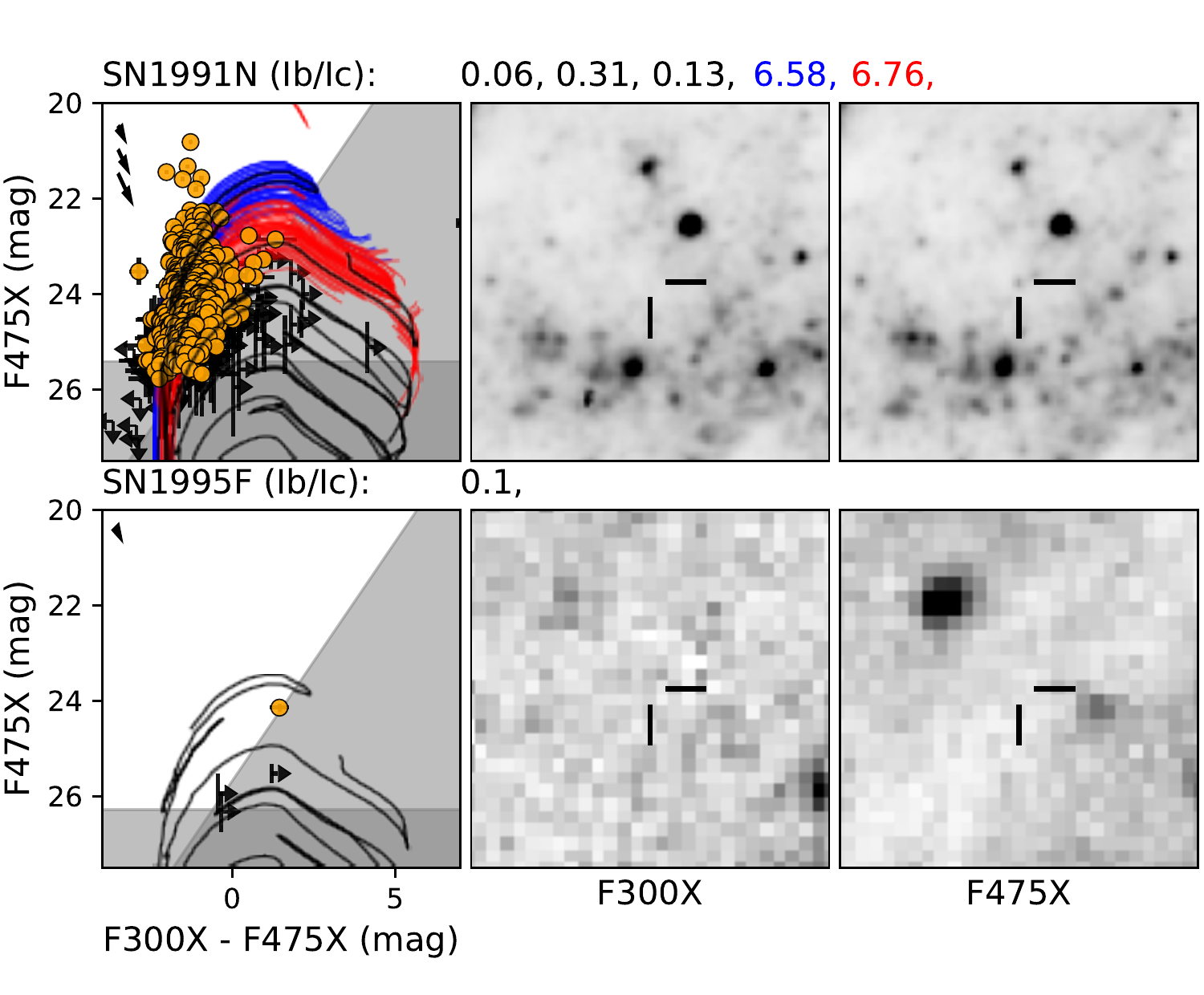}
\includegraphics[width=0.49\linewidth]{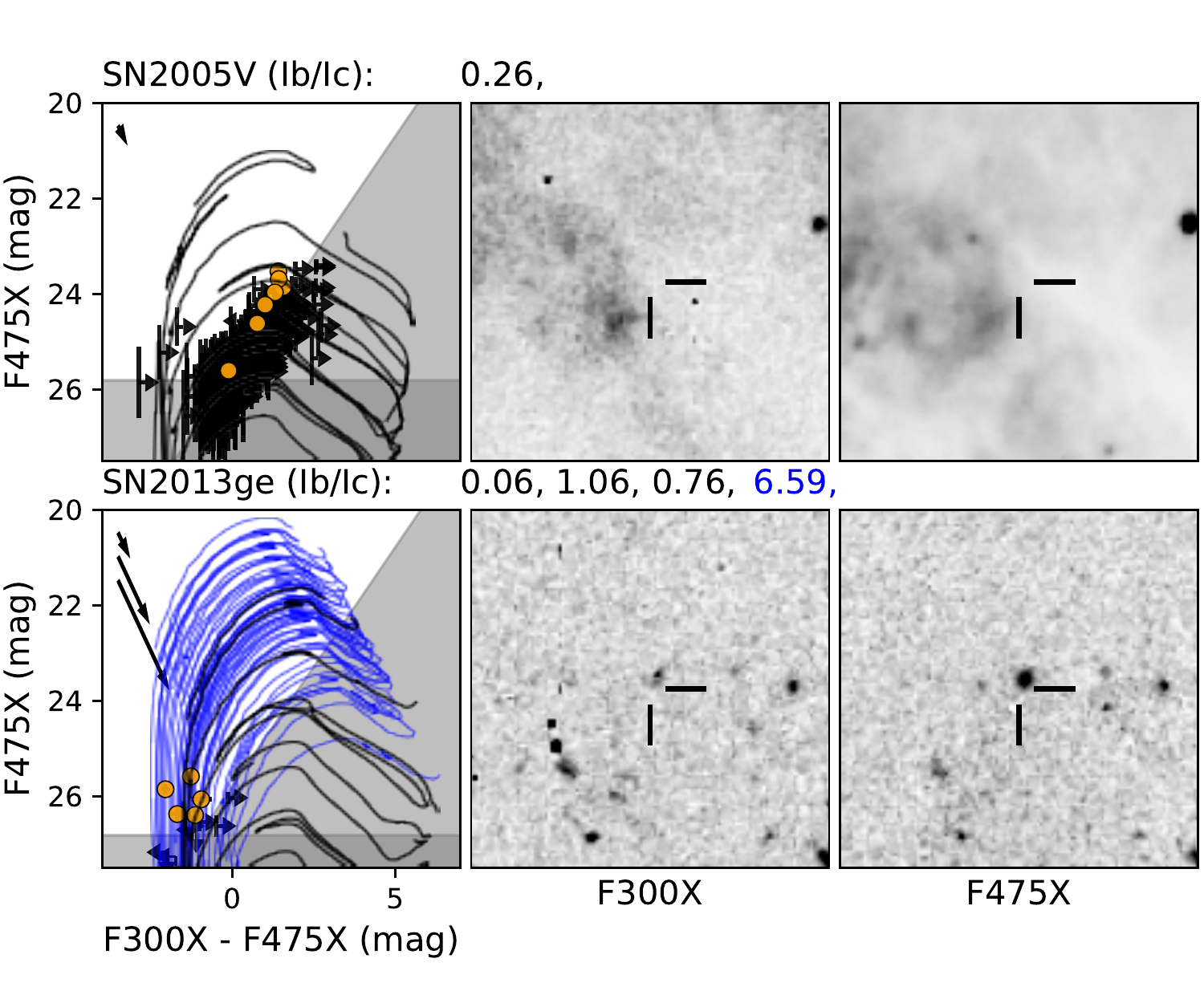}
\caption{Same as Fig.~\ref{IIb.fig} but for intermediate or ambiguous Type~Ib/Ic SNe.}
\label{IbIc.fig}
\includegraphics[width=0.49\linewidth]{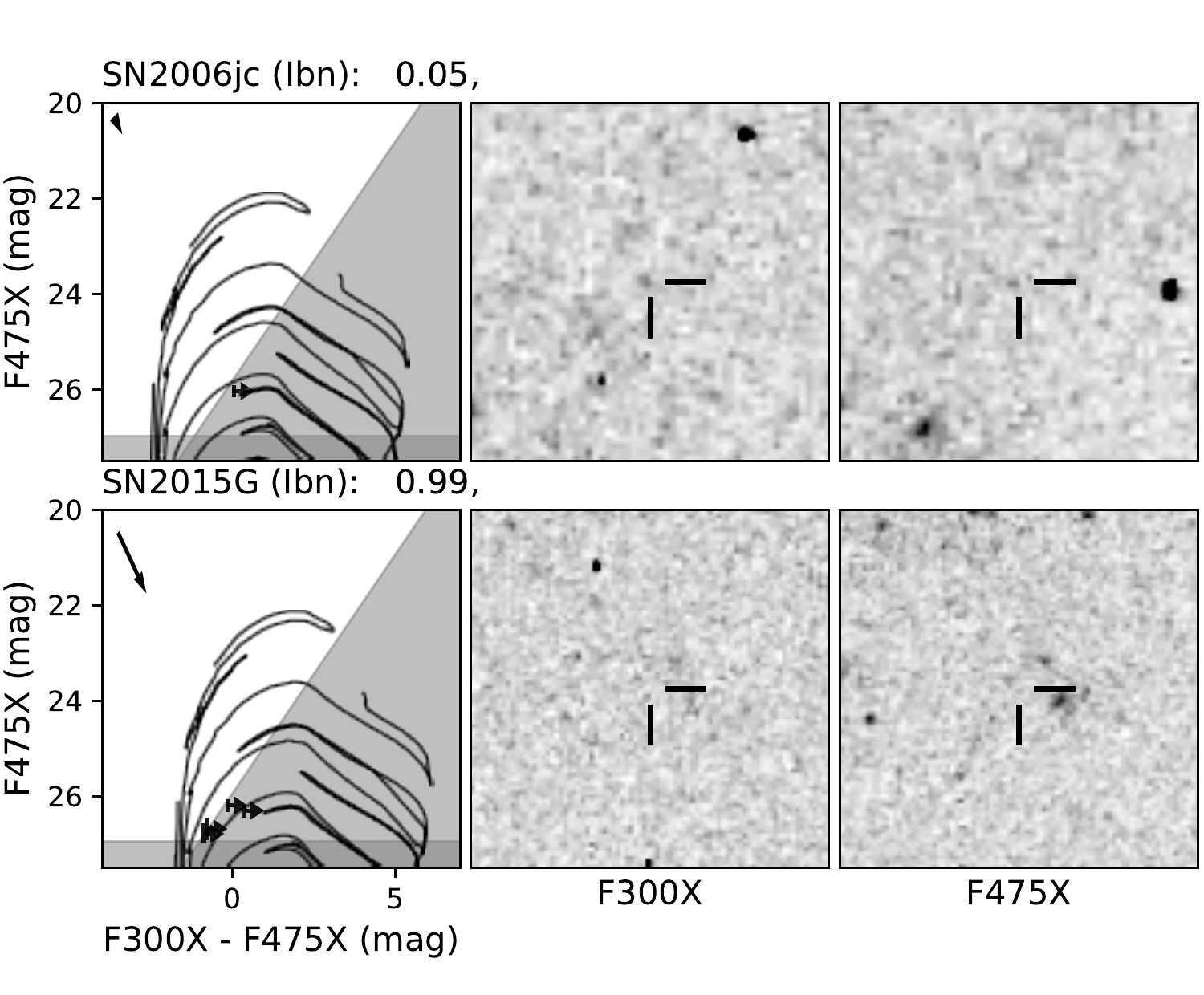}
\includegraphics[width=0.49\linewidth]{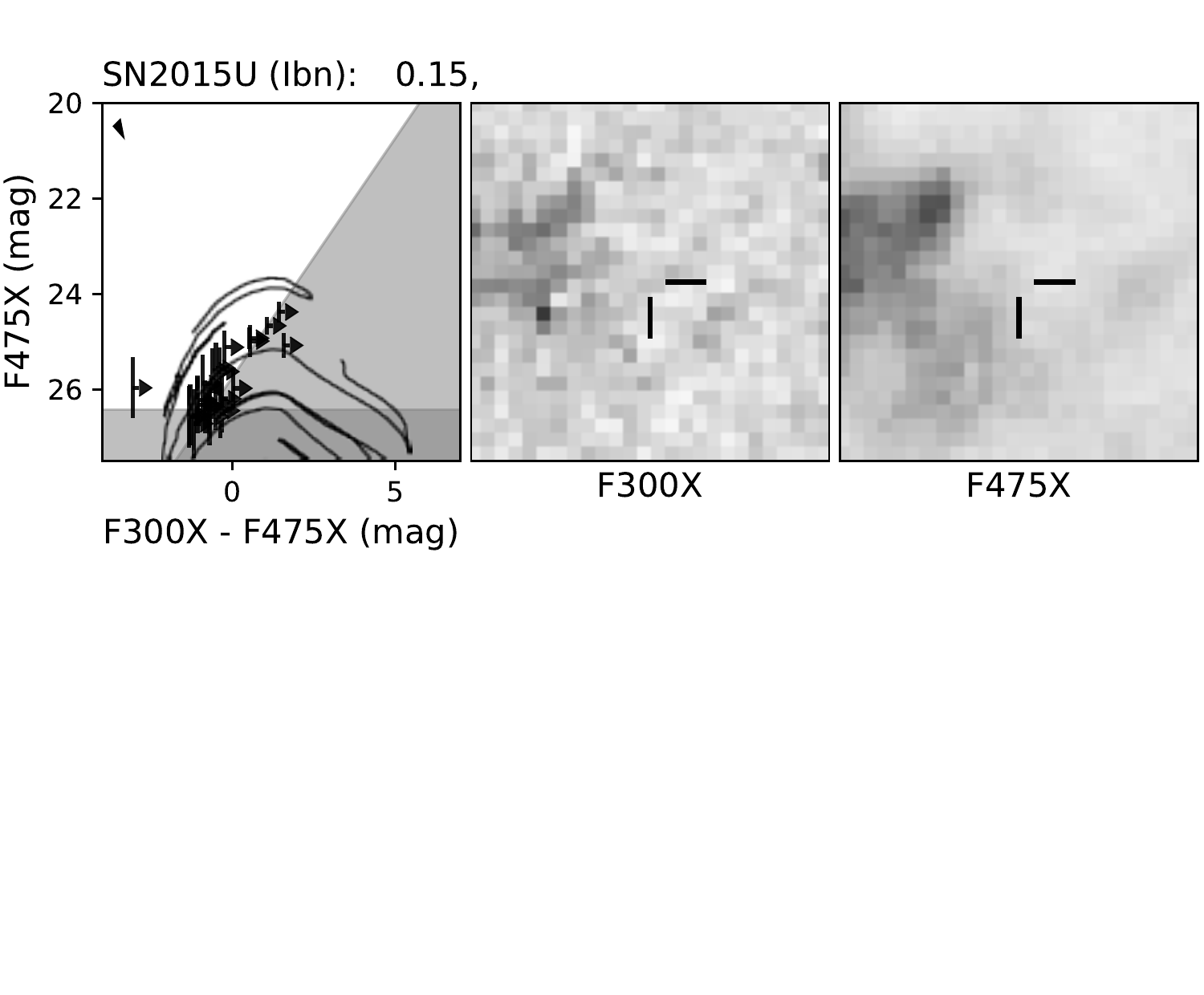}
\caption{Same as Fig.~\ref{IIb.fig} but for Type~Ibn SNe.}
\label{Ibn.fig}
\includegraphics[width=0.49\linewidth]{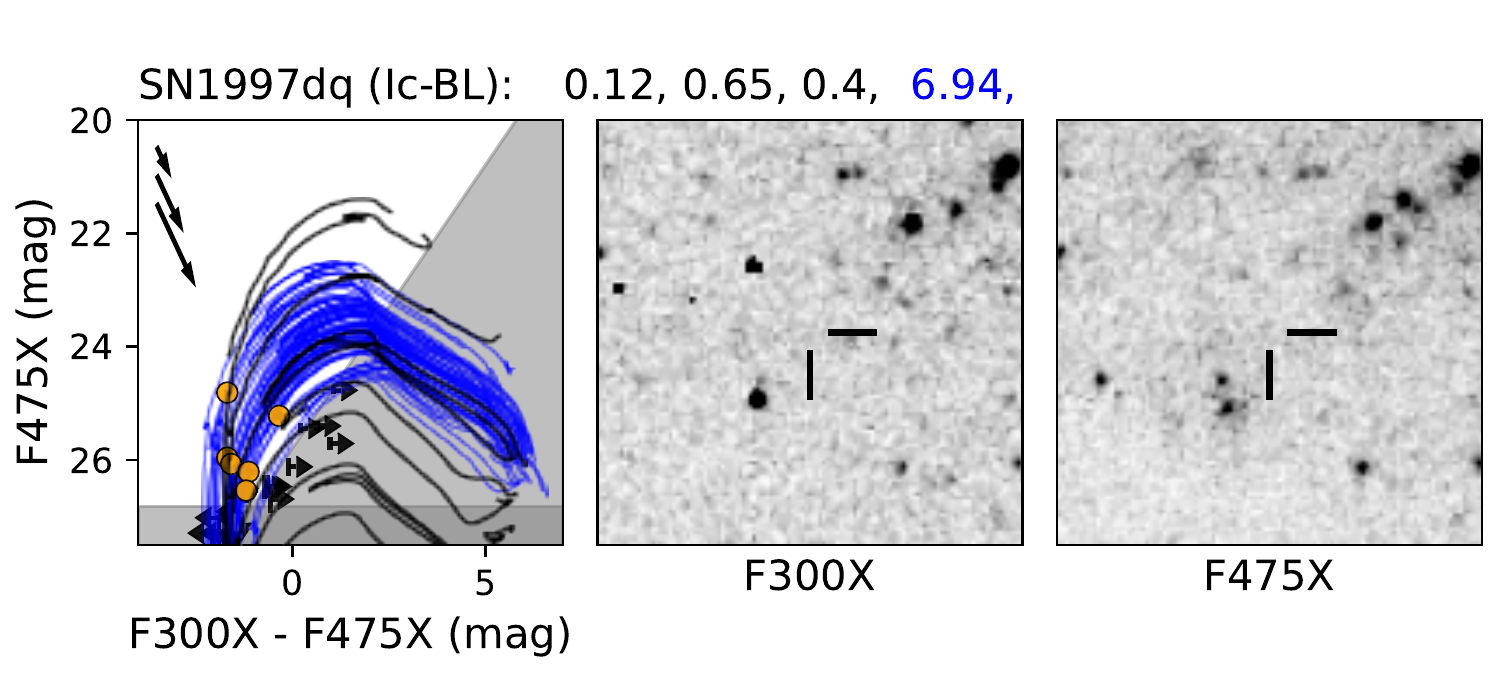}
\includegraphics[width=0.49\linewidth]{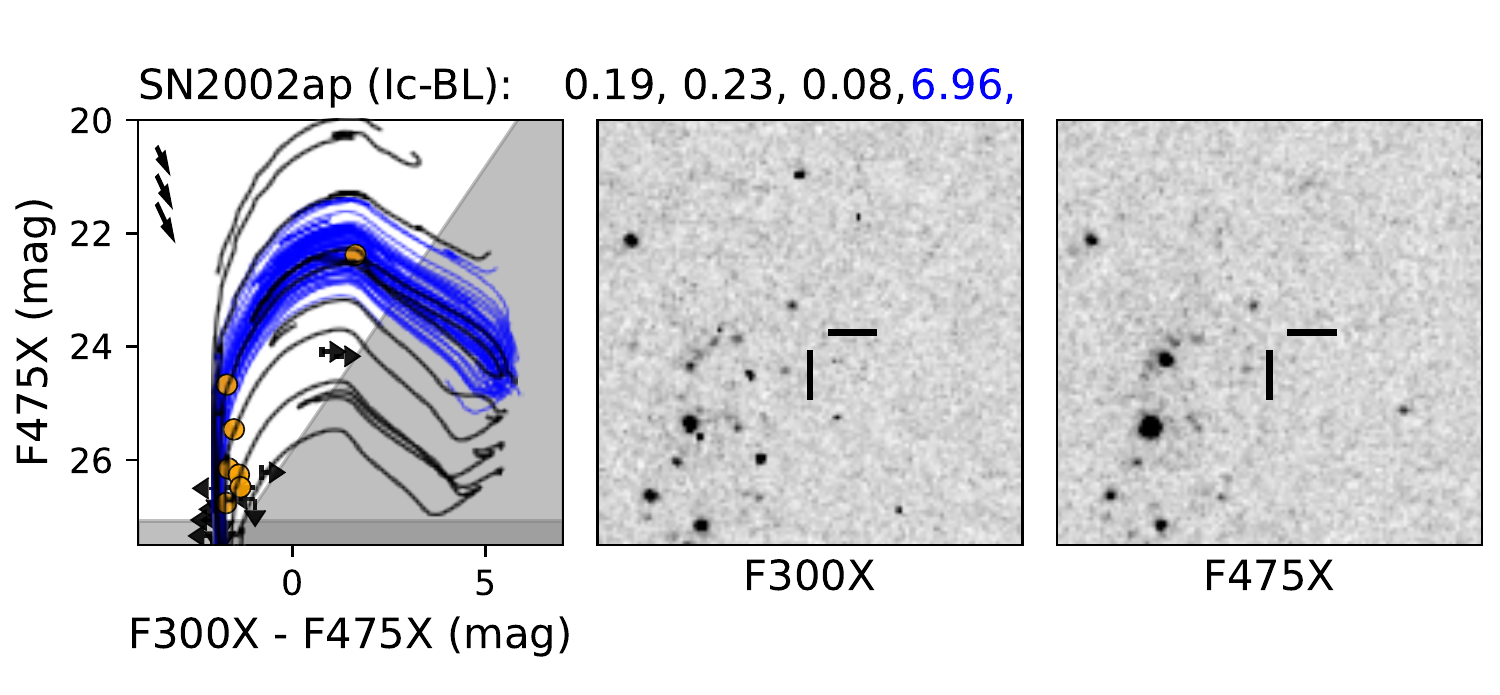}
\caption{Same as Fig.~\ref{IIb.fig} but for Type~Ic-BL SNe.}
\label{IcBL.fig}
\end{figure*}

\begin{figure}
\centering
\includegraphics[width=1\linewidth]{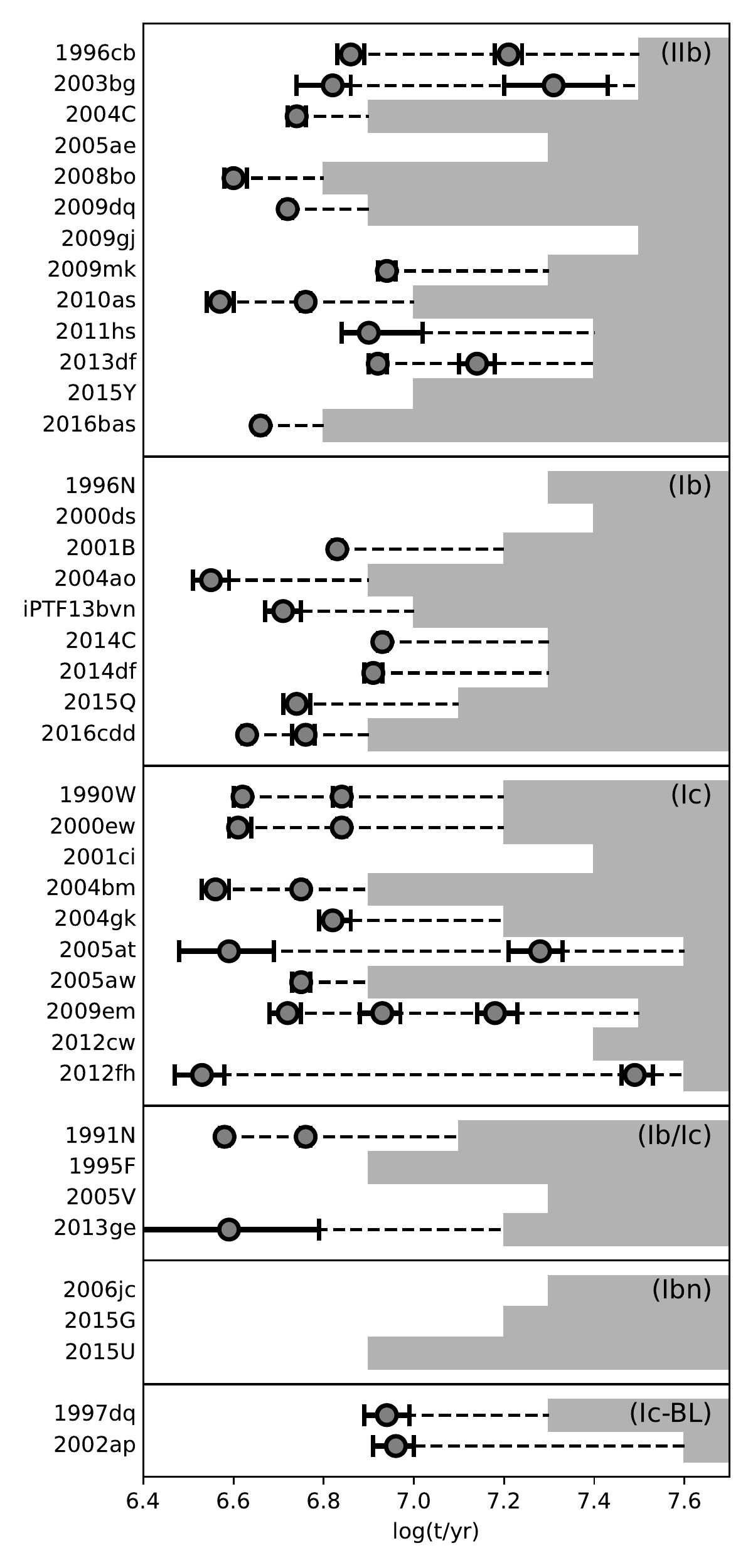}
\caption{Ages of the resolved stellar populations in the SN environments. The shaded regions show the age limits older than which a stellar population will fall below the detection limits.}
\label{pop.fig}
\end{figure}

\subsection{Target selection and observations}
\label{selection.sec}

We selected Type~IIb, Ib and Ic SNe from the IAU SN list\footnote{\url{http://www.cbat.eps.harvard.edu/lists/Supernovae.html}}, which occurred no earlier than the year of 1990, for the HST SNAP program ``\textit{A UV census of the sites of core-collapse SNe}" (PI: J. Maund; ID: 14762). The SNe were required to have a named host galaxy whose recession velocity corresponds to a Hubble-Flow distance of $\leq$25~Mpc (from the HyperLEDA database\footnote{\url{http://leda.univ-lyon1.fr}} and corrected for the infall toward Virgo;  \citealt{HyperLEDA.ref}). Only 9 SNe were excluded from the selection since they already had deep UV observations in the HST archive. Besides, we added some important SNe (which had been heavily observed as part of photometric and spectroscopic campaigns) to our sample with higher priority and relaxed distance constraints, such as the Type~Ibn and Type~Ic-BL SNe. This strategy ensures that the targets were selected in a uniform and unbiased way with a good coverage of SN types.

44 SNe were finally observed by the HST from 2016 Nov to 2018 May with the ultraviolet-visible (UVIS) channel of the Wide Field Camera~3 (WFC3). The extremely-wide UV filter F300X and optical filter F475X were used, with exposure times of 1200~s and 350~s, respectively,  for each target. Observations in each filter were divided into two individual exposures with a 2-point dithering. The limiting magnitudes were typically 25.7~mag in F300X and 26.6~mag in F475X (magnitudes are reported in the Vega system throughout this paper). The choice of the extremely-wide filter F300X and the long exposure times ensure that the young stellar populations can be probed down to very deep limits.

Out of the observed SNe, we later found SN~2006oz is actually a superluminous SN in a very distant dwarf galaxy \citep{sn2006oz.ref} and SN~2008ha is a peculiar Type~Iax SN whose nature is not yet clear \citep{sn2008ha.ref}; these two SNe were excluded in this work. SN~2005ar was also excluded since it is located at more than 100~Mpc away according to more recent distance measurements. The remaining 41 SESNE, listed in Table~\ref{sample.tab}, were finally used for the following analysis, including 13 IIb, 9 Ib, 10 Ic, 4 intermediate or ambiguous Ib/Ic, 3 Ibn and 2 Ic-BL.

Table~\ref{sample.tab} also provides the Galactic extinctions and the host galaxy inclinations and distances. Galactic extinctions are based on the estimates by \citet{Schlafly2011} and obtained from the NASA/IPAC IRSA Dust Extinction Service\footnote{\url{https://irsa.ipac.caltech.edu/applications/DUST}}. Host galaxy inclinations are from the HyperLEDA database. Distances to the SN host galaxies, when possible, are from the CosmicFlows project\footnote{\url{http://edd.ifa.hawaii.edu}} \citep{CosmicFlow2.ref, CosmicFlow3.ref}; for the other galaxies, we use Hubble-Flow distances adopting a Hubble constant of $H_0$~=~73.0~$\pm$~5~km~s$^{-1}$~Mpc$^{-1}$. Note that, although we aimed to select targets within 25~Mpc based on their recession velocities, some of the observed SNe have larger distances according to the more accurate CosmicFlows project.

We have not included any SESNe with archival HST data in order to ensure the uniformity in source selection and observations. The archival data are from various observing programs with different source selection criteria; therefore, including them would introduce some nonuniformity which itself is complicated and difficult to assess. For example, if the archival observations were aimed at studying active galactic nuclei, the SNe within the images would appear in the inner parts of the galaxies and those in the outer parts may fall outside the fields of view. In addition, the archival data had a range of filters and predominantly in the optical (there were only 9 SNe observed in the UV before this program, although the number has increased recently). By sticking with the extremely-wide F300X and F475X filters (to obtain high throughputs, which are rarely used by other observations) and the same exposure times for all targets, we try to acquire a uniform set of observations.

\subsection{Image processing and photometry}
\label{phot.sec}

Images processed by the standard reduction pipeline (\texttt{*\_flc.fits} and \texttt{*\_drc.fts}) were retrieved from the Mikulski Archive for Space Telescopes \footnote{\url{https://archive.stsci.edu/index.html}}. The images still contain a high level of cosmic-ray residuals; therefore, we re-drizzled them with \texttt{driz\_cr\_grow = 3} with the \textsc{astrodrizzle} package\footnote{\url{http://drizzlepac.stsci.edu/}}, which can clean the cosmic rays much more efficiently (all other parameters were unchanged).

We performed photometry with the \textsc{dolphot} package \citep{dolphot.ref} by fitting model point spread functions (PSFs) to the sources. The \textsc{dolphot} parameters \texttt{FitSky = 2} and \texttt{Img\_RAper = 3} were used, which perform better in crowded regions. We turned off aperture corrections with \texttt{ApCor = 0} since some images have very few stars for this purpose, causing large uncertainties (note that typical aperture corrections are only about several hundredths magnitude, which is much smaller than the random errors). \texttt{Force1 = 1} was used for better modeling of close stars that may be blended with each other. All other parameters were the same as those recommended in the \textsc{dolphot} user manual\footnote{\url{http://americano.dolphinsim.com/dolphot/dolphotWFC3.pdf}}.

A detection reported by \textsc{dolphot} is considered to be a good star if its quality parameters meet the following criteria:

(1) type of source, TYPE~= 1;

(2) signal-to-noise ratio, SNR~$\geq$~5;

(3) source sharpness, $-$0.5~$\leq$~SHARP~$\leq$~0.5;

(4) source crowding, CROWD~$\leq$~2;

(5) photometry quality flag, FLAG~$\leq$~3.

\noindent
For stars detected in only one filter, we used artificial stars, inserted at each source position, to estimate the detection limit in the other filter. An artificial star was considered to be successfully recovered if it fell within 1~pixel of the inserted position and its \textsc{dolphot} quality parameters met all the above criteria. The detection limits were derived independently for each star since the sky background and source crowding may vary across the field. We also used randomly positioned artificial stars to estimate additional photometric uncertainties induced by source crowding and imperfect sky subtraction.

We determined SN positions on the HST images with three different methods depending on the availability of supporting observations: (1) relative astrometry with previous HST observations, with typical uncertainties of $<$1~pixel; (2) relative astrometry with observations from other telescopes (e.g. Swift/UVOT and Spitzer/IRAC), with typical uncertainties of $\sim$3~pixel; and (3) using the reported SN R.A. and Dec. coordinates, with typical uncertainties of $\sim$1". Sources within a (projected) distance of 150~pc from the SN positions (excluding those within 3~pixels from the SNe) were finally selected for the following analysis (we did not correct for the host galaxies' inclinations). Consistent with the previous studies \citep[e.g.][]{Maund2017, Maund2018}, the size of the selection region was chosen to be compatible with the typical scales of star-forming complexes, i.e. fundamental cells of star formation within which the stars are born with similar ages \citep{Efremov1995}. The SN positional errors are not important compared with the defined area. Figures~\ref{IIb.fig}--\ref{IcBL.fig} show the images of the observed SN sites and the CMDs of the detected sources in their environments.

\subsection{Fitting of the resolved stellar populations}
\label{fitting.sec}

\subsubsection{Method}
\label{method.sec}

In order to derive the stellar ages, we used a hierarchical Bayesian approach to fit model stellar populations to the observed stars in the SN environments.  The method is described in detail in \citet{Maund2016} and \citet{Sun2021} and has been successfully applied in a number of studies (e.g., \citealt{Maund2017, Maund2018, Sun2020a, Sun2021, Sun2022a}). In brief, the model populations were simulated based on the \textsc{parsec} stellar isochrones (v1.2S; \citealt{parsec.ref}) and have a \citet{imf.ref} initial mass function, a 50\% binary fraction, and a flat distribution of primary-to-secondary mass ratio. In addition to the Galactic extinction, the internal extinctions from the host galaxy were assumed to have a Gaussian distribution $\mathcal{N}(A_V^{\rm int}, ({\rm d}A_V^{\rm int})^2)$ among the stars in each SN environment, where $A_V^{\rm int}$ is the mean value and ${\rm d}A_V^{\rm int}$ is the standard deviation; we used a flat prior for $A_V^{\rm int}$ and a logarithmic prior for ${\rm d}A_V^{\rm int}$ over the range of 0 $\leq$ log(${\rm d}A_V^{\rm int}$/0.01~mag) $\leq$ 2, penalizing large values. In each SN environment, the observed stars can be considered as arising from a mixture of model stellar populations with different ages and each model population corresponds to a short burst of star formation, in which the stellar log-ages follow a narrow Gaussian distribution with a standard deviation of 0.05~dex.

To determine the required number of model populations, we started the fitting with one age component, and added more age components if the fitting was significantly improved. We compared the fitting performances using the Bayes factor $K_{n+1, n}$ = ($Z_{n+1}$/$(n+1)!$)/($Z_n$/$n!$) and the criterion of \citet{bayes.ref}, where $Z_{n+1}$ and $Z_n$ are the Bayesian evidences for the fitting with $n+1$ and $n$ components [note that factors of $n!$ and $(n+1)!$ are needed to correct for overestimation since the labels for the model populations are interchangeable with each other]. We stopped adding more components if this did not improve the fitting significantly or if the number of components reached a maximum value of 3. In practice, all SNe, except the Type~Ic SN~2009em, require fewer than 3 age components to model the detected populations.

We did not perform any fitting when the SN environments contain no stars detected in both filters or only a few stars marginally detected above the detection limits. For each SN environment, we also estimated an age limit older than which a stellar population will fall below the detection limits; in doing this, we use the same extinction as derived from the population fitting, or simply assume zero internal extinction if no population fitting is performed for the SN (but this may overestimate the age limit).


\subsubsection{Caveats}
\label{caveats.sec}

At the distances of the analyzed SNe, the linear size of a pixel correspond to a few parsecs and the detected sources may or may not be single stars. In our population modelling, we have considered the possibilities of each source being a non-interacting binary system (Section~\ref{method.sec}; see also \citealt{Maund2016} and \citealt{Sun2021} for details), for which the composite fluxes are the simple sums of the primary and secondary stars' fluxes. This approach also accounts for the visual binaries composed of stars in chance alignment, thus source blending in crowded regions does not have a significant influence on our results (the visual binaries slightly increase the effective binary fraction, to which the population fitting is not very sensitive). However, this effect may still lead to an age underestimation in particular if the source lies in a region on the CMD where stars evolve very rapidly (and thus drives the fitting most sensitively). We have not considered the contribution from interacting binaries. The rejuvenated stars from binary mass transfer and merger may appear much younger \citep{Schneider2015}; therefore, our analysis tend to underestimate the stellar ages. It is, however, not trivial to model the interacting binary populations; this requires a detailed binary population synthesis and is beyond the scope of this work.

Star clusters that appear more extended than the PSF have been excluded by the selection criteria in Section~\ref{phot.sec} (extended sources have low sharpness values reported by \textsc{dolphot}). We cannot rule out the possibility that the stellar catalogues are still contaminated by very compact star clusters that are not spatially resolved due to their small angular sizes. In some SN environments some sources appear much brighter than the other sources and are more likely to be star clusters; they were manually removed from the fitting. For the ambiguous sources that are not easily identifiable by eye, our algorithm also automatically assigns lower weights to the ``outliers" in determining the model populations' parameters. For the low-mass star clusters whose brightnesses are often dominated by one or a few massive stars (e.g. the Orion Nebula Cluster), they can be fitted by our algorithm as if they are bright stars or binary systems. Proper quantitative assessment would require a detailed modelling of the star cluster population, which is beyond the scope of this work.

When available, we adopted metallicities for the model stellar populations that follow the reported values for the SNe. For those without metallicity estimates, we assume half-solar metallicity for the Type~Ic-BL SNe (which prefer low-metallicity environments; \citealt{Modjaz2020}) and solar metallicity for the other types. We ignored the metallicity uncertainties/dispersions and fixed their values in the fitting. This simplification greatly reduces the computing time. However, this may introduce some systematic errors since the stellar colours are metallicity-dependent.

Distances are fixed in our stellar population fitting. The typical uncertainties in distance moduli are 0.15--0.40~dex with a median value of 0.2~dex. By manually fitting stellar isochrones while keeping the extinction unchanged, we found that a 0.2-dex difference in the distance modulus roughly corresponds to a difference of 0.03~dex in stellar log-age. We also carried out a test by adding or subtracting 0.2~dex from the used distance modulus, and the derived ages changed by a negligible amount (in particular for the youngest population), since part of the distance difference has been absorbed by the internal extinction as a free parameter.

It was found that in some SN environments (e.g. SN~2004dg and SN~2012P; \citealt{Sun2021}) stars formed at different epochs may be located at different places along the line of sight and have different extinction values. For simplicity, however, our analysis assumes that $A_V^{\rm int}$ and ${\rm d}A_V^{\rm int}$ are independent of stellar ages. While this may introduce some systematic uncertainties, this simplification does not affect the youngest populations (which we shall focus on) that dominate the observed stars and drive the fitting in most circumstances. In the analysis we have used a standard extinction law with $R_V$ = 3.1 \citep{avlaw.ref} for both the Galactic and internal extinctions. We repeated the fitting with different values of $R_V$ = 2.0 and 5.0 for the internal extinction; although the derived extinctions may be different, the populations' ages change very little by $<$0.05~dex. Therefore, the reported ages are relatively insensitive to the choice of the reddening law. The fitting results are also unaffected by the uncertainties in the Galactic extinction, since any such errors will be absorbed by the internal extinction, which is a free parameter to be fitted.

We have not considered the stochastic sampling effect when the number of data points is very small. If the fitting is driven by only a few data points, the nominal uncertainty propagated from the photometric errors could be significantly underestimated. While it is difficult to make quantitative measurement, we caution this effect and provide the numbers of data points used in the fitting in the last column of Table~\ref{pop.tab}.

Lastly, we have assumed zero uncertainty in the adopted stellar isochrones. However, the evolution for young and massive stars is not well constrained and different sets of models could predict different evolutionary tracks \citep{Martins2013}. The model uncertainty is another source of error in our results.

\section{Results and discussions}
\label{results.sec}

\begin{figure}
\centering
\includegraphics[width=1\linewidth]{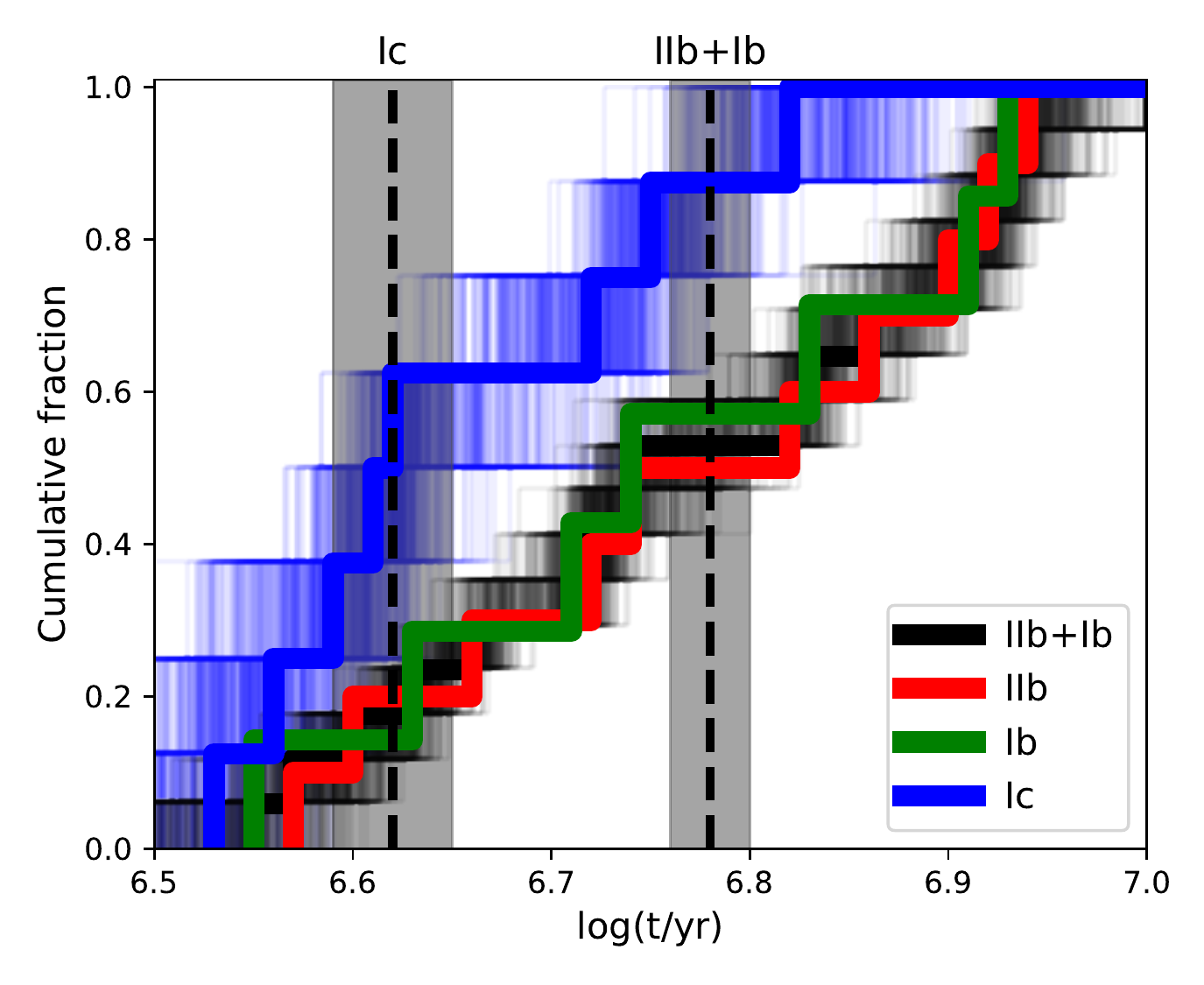} \\
\includegraphics[width=1\linewidth]{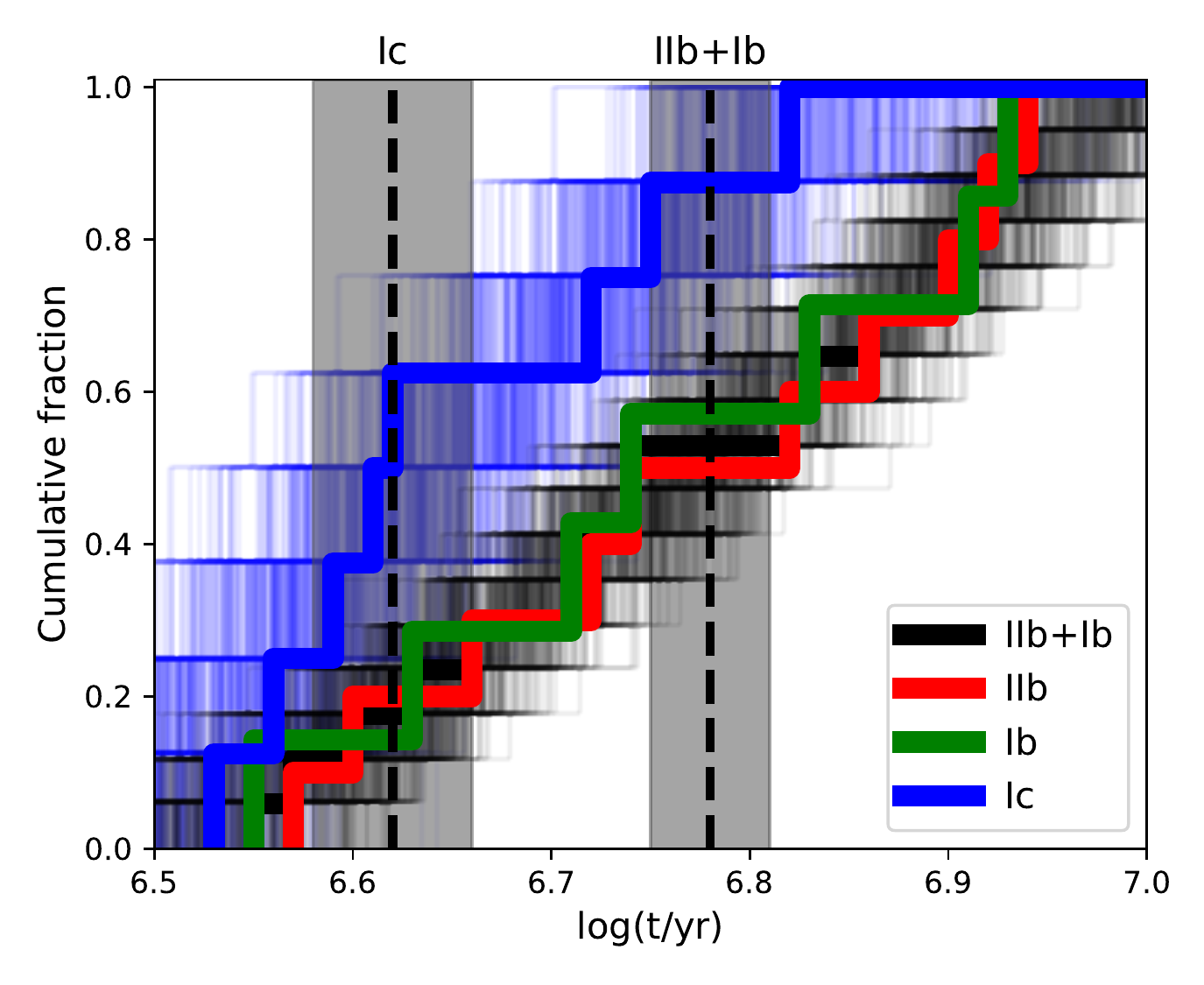}
\caption{Cumulative age distributions of the detected youngest stellar populations in the Types~IIb (red), Ib (green) and Ic (blue) SN environments. The total cumulative age distribution for the combined sample of Types~IIb+Ib SNe is shown in black. The thinner lines are 500 random realisations derived from the age uncertainties; in the upper panel, the uncertainties are propagated from the photometric uncertainties while in the lower panel, we conservatively assume an additional uncertainty of 0.05~dex from sources unaccounted for in our modelling (such as distance, extinction law, small number statistics, etc.). The median log-ages are shown by the vertical dashed lines with the shaded regions reflecting their uncertainties.}
\label{stat.fig}
\end{figure}

\begin{figure*}
\includegraphics[width=0.9\linewidth]{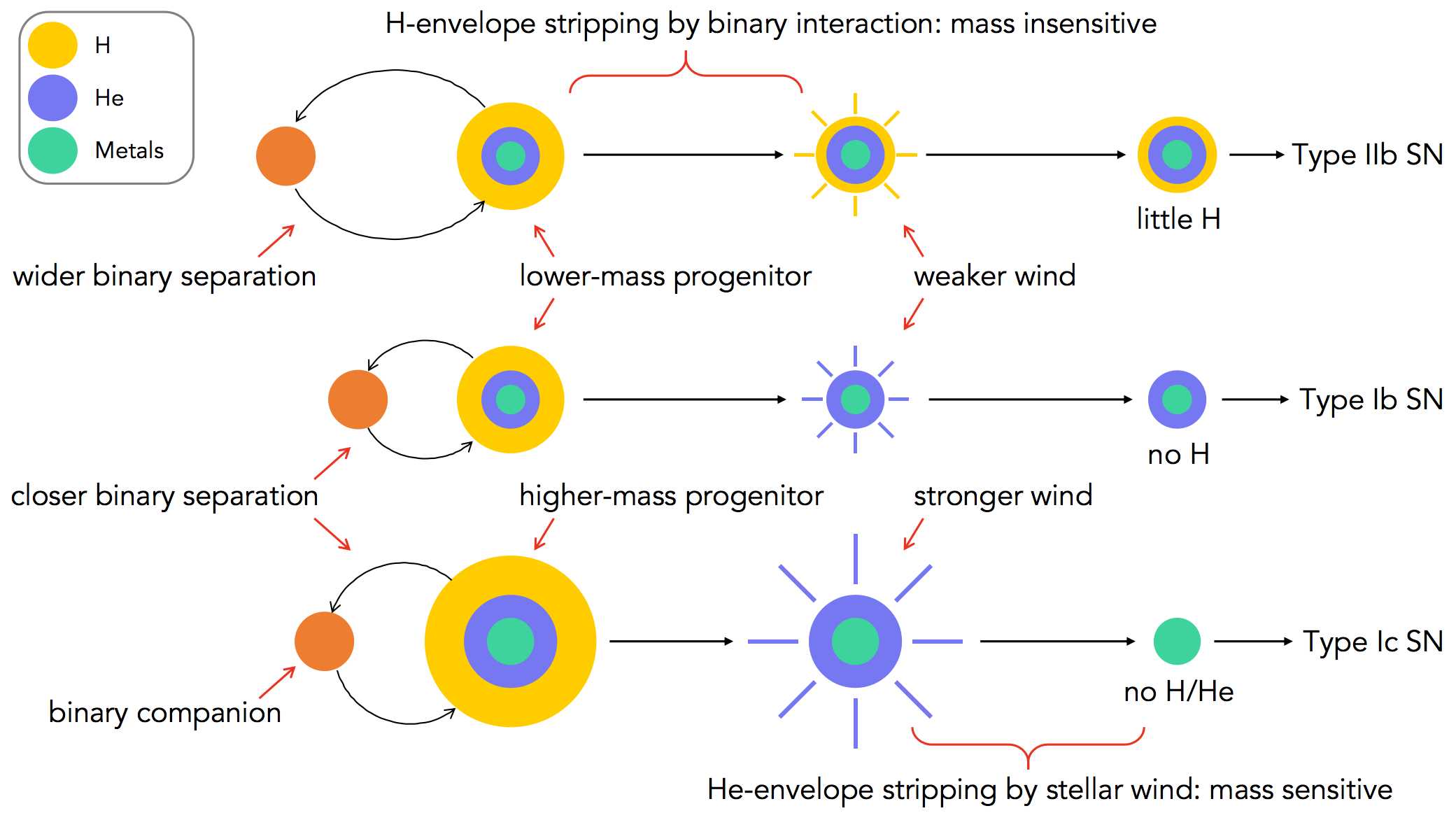}
\caption{A hybrid envelope-stripping mechanism that dominates the origins of Types~IIb, Ib and Ic SNe.}
\label{model.fig}
\end{figure*}

The population fitting results are listed in Table~\ref{pop.tab} and also displayed in Fig.~\ref{pop.fig}. The quoted uncertainties are propagated from the photometric errors and are estimated with the 68\% equal-tailed credible intervals (we also explore the effects of additional uncertainties in Section~\ref{normal.sec}); therefore, the uncertainties become larger when there is a significant degeneracy between age and extinction with double-peaked posterior probability distributions. The results show that very young stellar populations are present in the environments of a large fraction of the SESNe. $\sim$80\% of the observed Types~IIb, Ib and Ic SNe have stellar populations younger than 10~Myrs in their vicinity (10/13 Type~IIb,  7/9 Type~Ib, 8/10 Type~Ic, and 2/4 SNe that are of intermediate or ambiguous Type~Ib/Ic) and this fraction may be even higher if some SN environments contain young stars from residual star formation that have eluded our detection. For the Type~Ic SN~1997dq and SN~2002ap, young stellar populations are detected in their environments with ages of log($t$/yr)~= 6.94 and 6.96, respectively. In contrast, the Type~Ibn SN~2006jc and SN~2015G are located in very sparse areas without any obvious recent star formation, and SN~2015U does not have any resolved stars in its vicinity that are significantly detected in the UV filter.

It is worth noting that the age limits (older than which a stellar population will not be detected in both filters) are not the same for all targets and are strongly dependent on their distances and extinctions. Therefore, we shall characterise the age of each SN environment not with all the detected populations, but only with the youngest one, which is less affected by the inhomogeneous age limit.

The presence of very young stellar populations in the environment is often interpreted as the SNe having very massive progenitors with similarly young ages (e.g. WR stars). However, it should be cautioned that the SN progenitors may actually be much older and reside just in chance alignment. Also note that star formation bursts on small scales may appear correlated if they are controlled by physical processes on larger scales (e.g. spiral density waves that sweep up and compress the up-stream low-density gas and leave behind a trail of sequential star formation at the downstream, \citealt{Sun2021}; the feedback from massive stars which can trigger new episodes of star formation in their vicinity, \citealt{Koenig2008}; and the ubiquitous supersonic turbulence that drives hierarchical patterns in the spatial distribution of star formation, \citealt{Sun2017a, Sun2017b, Sun2018, Miller2022}). The relationship between a SN progenitor and the youngest stellar population in its environment may, therefore, be very complicated: the progenitor may (1) be a member of the youngest population and have the same age, (2) form at an earlier epoch of star formation \textit{unrelated} to the youngest population and aligned just by chance, or (3) arise from an older stellar population in the vicinity of the youngest population, whose formation is \textit{related} to each other.

Statistically, however, it is still reasonable to assume that SNe with more massive progenitors will have a higher probability to be associated with younger stellar populations in their environments (and vice versa; since the targets were uniformly selected, observed and analysed). Following this assumption, we discuss the environments and progenitors for the SESN types in the following subsections.

\subsection{Types~IIb, Ib and Ic}
\label{normal.sec}

Figure~\ref{stat.fig} compares the cumulative age distributions for the youngest stellar populations detected in the Types~IIb, Ib and Ic SN environments. The distributions are very similar between Type~IIb and Type~Ib SNe, while that for Type~Ic SNe is systematically much younger. We performed a two-sample Anderson-Darling test\footnote{Unlike the Kolmogorov-Smirnov test, which uses only the supremum distance between two cumulative distribution functions (CDFs), the Anderson-Darling test uses the sum of all squared distances between two CDFs to test the null hypothesis; therefore, it is more sensitive to CDF differences in the tails.} between Type~Ic and Types~IIb+Ib; the null hypothesis that the two samples are drawn from the same distribution can be rejected at a 6\% significance level. The median log-ages of the detected youngest populations are log($t$/yr)~= 6.62 $\pm$~0.03 for Type~Ic SNe and 6.78 $\pm$~0.02 for Types~Ib+IIb SNe. This suggests that the Type~Ic SN progenitors are on average younger and more massive than those of Type~IIb and Type~Ib SNe, while the Type~IIb and Type~Ib SNe have on average indistinguishable progenitor masses.

In the upper panel of Fig.~\ref{stat.fig}, the age uncertainties are propagated only from the photometric uncertainties. As mentioned in Section~\ref{caveats.sec}, however, there could be other error sources not included in our population fitting. While the interacting binaries and ambiguous star clusters may lead to an age underestimation, their effects are similar for different SN types since our analysis is performed in a uniform way. For the other error sources (distances, extinction law, small number statistics, etc.), we conservatively assume that they contribute an additional uncertainty of 0.05~dex to the age determination (see the discussion in Section~\ref{caveats.sec}). The updated age distributions are displayed in the lower panel of Fig.~\ref{stat.fig}; while the distributions become broader, the difference between Type~Ic and Types~IIb+Ib SNe is still significant. The uncertainties of the median log-ages increase by 0.01~dex.

We also use an alternative method without isochrone fitting to qualitatively check the environmental difference for the different SN types; if the Type~Ic SNe tend to be associated with younger stellar populations, their environments will have a larger number of bright stars than the Type~IIb and Type~Ib SNe. We calculate the absolute magnitudes of the surrounding stellar populations and correct their Galactic extinctions; apart from a few SNe with large distances and/or high extinctions (Type~IIb SN~2010as, SN~2015Y and SN2016bas; Type~Ib SN~2016cdd; and Type~Ic~SN~2005aw), the detections are complete down to limits of $m_{F300X}$ = $-$7.5~mag and $m_{F475X}$ = $-$6.5~mag. The Type~IIb SN~2009dq is also excluded since its environment contains many sources that are too bright to be single/binary stars but are more likely to be star clusters (Fig.~\ref{IIb.fig}). We find that the Types~IIb/Ib/Ic SN environments have on average 5/8/15 stars brighter than $m_{F300X}$ = $-$7.5~mag, and 7/5/17 stars brighter than $m_{F475X}$ = $-$6.5~mag, respectively. This is consistent with the previous finding that Type~Ic SN environments are on average younger than the Type~IIb and Type~Ib SNe. In this analysis we have not corrected for the internal extinctions within the host galaxies, but the stellar population fitting shows that Type~Ic SNe tend to have higher internal extinctions than the Type~IIb and Type~Ib SNe (Table~\ref{pop.tab}), so the number of bright stars could potentially be even larger for the Type~Ic SNe. All these results suggest that the Type~Ic SN environments and progenitors are significantly different from those of Type~IIb and Type~Ib SNe.

\subsubsection{A hybrid envelope-stripping mechanism?}
\label{model.sec}

\citet{Fang2019} analysed the late-time nebular spectra of 13 Type~IIb, 16 Type~Ib and 16 Type~Ic SNe. They found that Type~Ic SNe have larger [O~\textsc{i}]/[Ca~\textsc{ii}] line ratios than Type~IIb and Type~Ib SNe, while those for Type~IIb and Type~Ib SNe are very similar to each other.  This result is later justified by their more recent work \citep{Fang2022} based on a larger sample. The [O~\textsc{i}]/[Ca~\textsc{ii}] line ratio is a good indicator of the final carbon-oxygen core mass, which is higher for more massive SN progenitors. Therefore, they have reached the same conclusion as we draw from the above environmental analysis.

Our result therefore supports the hybrid envelope-stripping model proposed by \citet[][see also the schematic diagram in Fig.~\ref{model.fig}]{Fang2019} for the origins of SESNe. The indistinguishable progenitor masses between Type~IIb and Type~Ib SNe suggest that the hydrogen envelope is stripped by a mass-insensitive process, and binary interaction is the best candidate without major difficulties. The outcome of binary interaction is strongly dependant on the initial binary separation: a star in a relatively wide binary may end up with a small residual hydrogen envelope and explode as a Type~IIb SN, while a star of the same initial mass in a closer binary can completely lose its hydrogen envelope and explode as a Type~Ib SN \citep[e.g.][]{Claeys2011}. The effect of binary interaction depends weakly on the progenitor mass, and we note that a mass- and metallicity-dependant stellar wind may remove the residual hydrogen envelope left after the binary interaction \citep[e.g.][]{Yoon2017}.

The fact that Type~Ic SNe appear to favour higher progenitor masses than Type~IIb and Type~Ib SNe suggest that the stripping of the helium envelope is caused by a mass-sensitive process. This may be the stellar wind of the hydrogen-poor progenitor in the post-binary interaction phase, which is stronger for more massive stars (\citealt{Vink2017}; note that binary interaction alone is very difficult to remove the entire helium envelope in self-consistent models; \citealt{handbook.ref}). It is also possible that a very massive progenitor can lose all of its hydrogen- and helium-rich layers solely via its own stellar wind without the aid of binary interaction. Some Type~Ic SNe may still have modest He envelopes that are not completely stripped, but their helium lines are not excited due to, e.g., the weak mixing between helium and radioactive elements \citep{Dessart2012}.

The differences between Types IIb+Ib and Type Ic SN progenitors found by \citet{Fang2019} and this work are only \textit{statistically} meaningful instead of being specific about individual progenitors. It is possible that a Type~IIb or Type~Ib SN progenitor could be more massive than that for a Type~Ic SN. Similarly, the proposed hybrid envelope-stripping model may just account for most, instead of all, of the observed SN populations, and there could be many other progenitor channels for the origins of SESNe \citep[e.g.][]{Yoon2017b, Yoon2017, Hirai2020}.

\subsubsection{Comparison with previous works}
\label{comparison.sec}

The environments of SESNe have been studied in a number of previous works with a variety of methods. For example, \citet{Galbany2018} and \citet{K2018} used the H$\alpha$ equivalent width to estimate the ages of the ionising stellar populations, and \citet{Anderson2012} and \citet{Kangas2013} analysed the SNe' spatial correlation with H$\alpha$ emission (and other tracers of star formation). \citet{Kangas2013} showed that Type~Ic SNe are most strongly correlated with recent star formation, while the difference between Type~II and Type~Ib SNe is insignificant (consistent with our results). On the other hand, however, some of the works found a trend that, following Types~IIb $\rightarrow$ Ib $\rightarrow$ Ic, the SNe occur in progressively younger environments \citep[e.g.][]{Anderson2012, K2018}.

We caution that the use of the H$\alpha$ equivalent width as a stellar age indicator may suffer from large uncertainties. Its value is very sensitive to the covering factor, the older stellar populations along the same line of sight (which contribute to the stellar continuum), and whether the effect of interacting binaries is taken into account. Recent works \citep[e.g.][]{Xiao2018, Schady2019, Xiao2019} found that the derived stellar ages can be very different if these effects were or were not considered; \citet{Sun2021} showed that accurate stellar ages can only be obtained by a detailed photoionisation calculation and a full modelling of the nebular spectral lines. Large uncertainties may also affect the spatial correlation analysis. As pointed out by \citet{Crowther2013}, the giant H~\textsc{ii} regions (rather than isolated and compact ones) in the vicinity of most SN sites can undergo complicated and multi-generation evolution, and the observed spatial association provides only weak constraints on the SN progenitor masses. Moreover, some SNe apparently associated with H~\textsc{ii} regions on the low-spatial resolution images may actually have an offset between each other \citep[e.g.][]{Maund2016}.

Studying  the resolved stellar populations allows a direct age estimate of the SN environments. \citet{Maund2018} analysed 23 SESNe and found the characteristic ages for the environmental populations (approximately the mean ages for all the detected populations for a given SN type, not just the youngest population in the environment) are log($t$/yr)~= 7.20, 7.05, and 6.57 for Types~IIb, Ib and Ic SNe, respectively. However, that work was based on a heterogeneous dataset and the effect of selection bias is very difficult to estimate. In addition, many of the analysed SNe lack UV observations; since the optical colour is very insensitive to stellar age, the age determination may therefore have relatively large uncertainties, especially for the youngest populations of only a few Myrs.

In our analysis, the age distributions of the youngest environmental populations are almost identical for the Type~IIb and Type~Ib SNe (Fig.~\ref{stat.fig}). We cannot rule out the possibility that this is due to the measurement uncertainties and/or the limited sample size. Observations of larger samples are required to further confirm this result.

\subsection{Type~Ibn}
\label{ibn.sec}

Type~Ibn SNe are those which exhibit narrow helium lines with little evidence for hydrogen. The narrow helium lines are a signature of strong interaction between the SN ejecta and helium-rich CSM. The CSM was formed via eruptive mass loss of their progenitors shortly before their explosions \citep{Smith2017}. Early studies proposed that their progenitors are massive WR stars stripped by their powerful winds \citep[e.g.][]{sn2006jc.ref1, sn2006jc.ref2}. More recently, \citet{sn2015g.ref} ruled out a WR progenitor for SN~2015G by examining its pre-explosion images. \citet{Sun2020a} found a progenitor initial mass of only $\lesssim$12~$M_\odot$ for SN~2006jc by analysing its surviving binary companion. Modelling of the SN spectra \citep{Dessart2022} showed that Type~Ibn SN progenitors have final masses of only $\lesssim$5~$M_\odot$, much lower than those for WR stars.

For the Type~Ibn SN~2006jc and SN~2015G, our deep imaging did not detect any obvious stellar populations in their environments younger than log($t$/yr)~= 7.3 and 7.2, respectively (Table~\ref{pop.tab} and Fig.~\ref{pop.fig}). The age limits were derived by assuming zero internal extinctions within their host galaxies. If we adopt the extinction estimates by \citet{sn2006jc.ref2} and \citet{sn2015g.ref}, the age limit changes to log($t$/yr)~= 7.1 for SN~2015G while that for SN~2006jc is almost unchanged. The age limits correspond to progenitor masses of $M_{\rm ini}$~$\lesssim$ 11.9~$M_\odot$ for SN~2006jc and $M_{\rm ini}$~$\lesssim$ 13.8~$M_\odot$ (assuming zero internal extinction) or 16.3~$M_\odot$ (using the extinction estimate from \citealt{sn2015g.ref}) for SN~2015G. These constraints argue against massive WR progenitors, whose lifetimes are only a few Myrs \footnote{Recently, \citealt{Beasor2021} derived an older age of $\sim$10~Myr for the Galactic cluster Westerlund~1, which host dozens of WR stars, from its cool supergiant population. If the WR stars have the same age as the cool supergiants, the age and initial mass limits for WR stars may have to be soften to some extent, possibly owing to their binary evolutionary channel. However, we also note that Westerlund~1 was formed in an extended period of star formation of several Myrs, so the WR stars in this cluster may still have very young ages consistent with previous constraints \citep{Crowther2007}.}. Their progenitors are most likely stripped via binary interaction, which then underwent eruptive mass loss shortly before their core collapse. These two processes led to the stripping of their hydrogen envelopes and the formation of the dense helium-rich CSM. Therefore, this result confirms that violent eruptions, often seen in very massive ($>$25--30~$M_\odot$) luminous blue variables, can also occur in much lower-mass stars stripped in binaries (\citealt{Fuller2018, Wu2022}; see also \citealt{Sun2020a} for a detailed discussion).

For the more distant SN~2015U, the age limit [log($t$/yr)~$>$ 6.9] is not sufficient to robustly reject a WR progenitor. This limit is likely to be overestimated since we assumed zero internal extinction but SN~2015U seems to be in a dusty area. Moreover, there is a UV blob in its vicinity, which could be an unresolved young star-forming complex. Therefore, the environment of SN~2015U is rather ambiguous.


\subsection{Type~Ic-BL}
\label{icbl.sec}

Type~Ic-BL SNe are characterised by the broad lines in their spectra, indicating extremely fast expansion velocities. Their explosion energies can reach $\sim$10$^{52}$~erg, almost an order of magnitude higher than the normal core-collapse SNe. \citet{Lyman2016} found relatively low ejecta masses for almost all Type~Ic-BL SNe in their sample, suggesting moderately massive progenitors of $M_{\rm init}$~= 8--20~$M_\odot$ stripped in binaries. In contrast, \citet{Taddia2019} showed that at least 21\% of the Type~Ic-BL SNe have very large ejecta masses of $>$5.5~$M_\odot$ and should arise from the more massive WR progenitors with $M_{\rm ini} >$ 28~$M_\odot$.

In our sample, the Type~Ic-BL SN~1997dq was poorly studied previously \citep[see, however,][]{Mazzali2004} and SN~2002ap has received far more attention. For SN~2002ap, the light curve analysis of \citet{Mazzali2002} derived a carbon-oxygen core mass of 5~$M_\odot$, corresponding to an initial mass of 20--25~$M_\odot$ for the progenitor. \citet{Crockett2007} ruled out any single progenitors that have not evolved into a WR phase based on detection limits on the pre-explosion images; they suggested that the progenitor, if it were indeed a single star, should have an initial mass of at least 30--40~$M_\odot$ and a higher-than-standard mass-loss rate. More recently, \citet{Zapartas2017} constrained the progenitor initial mass to be $\lesssim$23~$M_\odot$ with binary population synthesis.

\citet{Maund2018} detected only a few stars in the environment of SN~2002ap based on optical images acquired by the High Resolution Channel (HRC) of HST's Advanced Camera for Surveys (ACS). Our new images, which use the extremely wide F300X and F475X filters and reach deeper detection limits, clearly reveals a young and resolved stellar population in the vicinity of SN~2002ap (Fig.~\ref{IcBL.fig}). This is consistent with the deep UV images acquired by \citet[][their Fig.~1]{Zapartas2017} in the F275W and F336W filters with exposure times of 2778~s in each band. The derived age [log($t$/yr)~= 6.96$^{+0.04}_{-0.05}$] corresponds to a progenitor initial mass of $M_{\rm ini}$~= 21.6$^{+2.5}_{-1.8}$~$M_\odot$, assuming the progenitor is a member of  and has the same age as the detected population. This value is consistent with the previous estimates as mentioned above. For SN~1997dq, the detected environmental population suggests a progenitor mass of $M_{\rm ini}$~= 22.5$^{+2.8}_{-2.3}$~$M_\odot$, very similar to that of SN~2002ap. It is also possible, however, that the progenitor may have a lower initial mass if it belongs to an undetected older stellar population. These results are inconsistent with massive WR progenitors and agree with a binary progenitor channel (but the envelope stripping may still be aided by its own stellar wind). Note that Type~Ic-BL SNe prefer low-metallicity environments \citep{Modjaz2020}, where stars have weaker winds and the initial mass threshold can be as high as $\sim$30~$M_\odot$ for a single star to evolve into the WR phase \citep{Crowther2007}.

\section{Summary and conclusions}
\label{summary.sec}

This paper reports an environmental analysis of 41 SESNe. The targets were selected and observed uniformly and unbiasedly, and deep UV-optical imaging was conducted by the HST to probe the young stars in the SN environments. We used a hierarchical Bayesian approach to fit resolved stellar populations on the CMDs in order to derive their ages and extinctions. The high sensitivity of UV-optical colour to stellar age/extinction allows us to make very precise measurements, with typical uncertainties smaller than 0.1~dex in log-age and 0.1~mag in extinction in most cases.

As many as 80\% of the observed Types~IIb, Ib and Ic SNe have stellar populations younger than 10~Myrs in their environments. The age distributions for the youngest environmental populations for Type~IIb and Type~Ib SNe are very similar to each other, while that for Type~Ic SNe is systematically much younger. This suggests that Type~Ic SNe have more massive progenitors than Type~IIb and Type~Ib SNe; in contrast, no compelling evidence is found for any significant difference between Type~IIb and Type~Ib SN progenitor masses.

This result supports a hybrid model of envelope-stripping for the origins of SESNe (as proposed by \citealt{Fang2019}). In this scenario, the stripping of the hydrogen envelope is mostly caused by binary interaction as a mass-insensitive process, while that for the helium envelope could be due to a mass-sensitive mechanism, such as the wind of the hydrogen-poor progenitor at the post-binary interaction phase.

For the Type~Ibn SNe, our deep imaging excludes any young stellar populations in the environments of SN~2006jc and SN~2015G that correspond to massive WR progenitors. They most likely arise from moderately massive ($\lesssim$11.9 and $\lesssim$16.3~$M_\odot$, respectively) stars stripped in interacting binaries. For the more distant SN~2015U, however, the age limit is not sufficient to make any useful constraints. This result confirms that the very violent eruptions, often seen in the very massive luminous blue variables, can also occur in much lower-mass stars stripped in binaries, forming dense and helium-rich CSM shortly before their final explosion.

For the Type~Ic-BL SNe, we detect young stellar populations in the vicinity of both SN~1997dq and SN~2002ap. Their progenitors would have initial masses of $M_{\rm ini}$~= 22.5$^{+2.8}_{-2.3}$~$M_\odot$ and 21.6$^{+2.5}_{-1.8}$~$M_\odot$, respectively, if they are coeval with the detected populations. This result also argues against WR progenitors for both SNe and favours an interacting binary progenitor channel possibly aided by their own stellar winds.

\section*{Acknowledgements}

The authors are very grateful to Prof. Keiichi Maeda, Mr. Qiliang Fang and the anonymous referee for their very helpful comments on this paper. Research of N-CS, JRM and PAC are funded by the Science and Technology Facilities Council through grant ST/V000853/1. This paper is based on observations made with the NASA/ESA Hubble Space Telescope.

\section*{Data availability}
Data used in this work are all publicly available from the Mikulski Archive for Space Telescope (\url{https://archive.stsci.edu}).

\bsp	
\label{lastpage}

\begin{thebibliography}{99}

\bibitem[\protect\citeauthoryear{Aldering, Humphreys, \& Richmond}{1994}]{Aldering1994} Aldering G., Humphreys R.~M., Richmond M., 1994, AJ, 107, 662

\bibitem[\protect\citeauthoryear{Anderson et al.}{2012}]{Anderson2012} Anderson J.~P., Habergham S.~M., James P.~A., Hamuy M., 2012, MNRAS, 424, 1372


\bibitem[\protect\citeauthoryear{Beasor et al.}{2021}]{Beasor2021} Beasor E.~R., Davies B., Smith N., Gehrz R.~D., Figer D.~F., 2021, ApJ, 912, 16

\bibitem[\protect\citeauthoryear{Benvenuto \& Bersten}{2017}]{handbook.ref}Benvenuto O.~G., \& Bersten M.~C., 2017, Handbook of Supernovae. Springer International Publishing AG, Cham, p. 403

\bibitem[\protect\citeauthoryear{Beswick et al.}{2005}]{sn2005v.ref1} Beswick R.~J., Fenech D., Thrall H., Argo M.~K., Muxlow T.~W.~B., Pedlar A., 2005, IAUC, 8572

\bibitem[\protect\citeauthoryear{Bressan et al.}{2012}]{parsec.ref} Bressan A., Marigo P., Girardi L., Salas    nich B., Dal Cero C., Rubele S., Nanni A., 2012, MNRAS, 427, 127

\bibitem[\protect\citeauthoryear{Cao et al.}{2013}]{iptf13bvn.ref} Cao Y., Kasliwal M.~M., Arcavi I., Horesh A., Hancock P., Valenti S., Cenko S.~B., et al., 2013, ApJL, 775, L7

\bibitem[\protect\citeauthoryear{Childress et al.}{2015}]{sn2015y.ref} Childress M., Scalzo R., Yuan F., Zhang B., Ruiter A., Seitenzahl I., Schmidt B., et al., 2015, ATel, 7368

\bibitem[\protect\citeauthoryear{Chornock \& Filippenko}{2001}]{sn2001b.ref} Chornock R., Filippenko A.~V., 2001, IAUC, 7577

\bibitem[\protect\citeauthoryear{Chornock \& Berger}{2009}]{sn2009mk.ref} Chornock R., Berger E., 2009, CBET, 2086

\bibitem[\protect\citeauthoryear{Claeys et al.}{2011}]{Claeys2011} Claeys J.~S.~W., de Mink S.~E., Pols O.~R., Eldridge J.~J., Baes M., 2011, A\&A, 528, A131

\bibitem[\protect\citeauthoryear{Crockett et al.}{2007}]{Crockett2007} Crockett R.~M., Smartt S.~J., Eldridge J.~J., Mattila S., Young D.~R., Pastorello A., Maund J.~R., et al., 2007, MNRAS, 381, 835

\bibitem[\protect\citeauthoryear{Crockett et al.}{2008}]{Crockett2008} Crockett R.~M., Eldridge J.~J., Smartt S.~J., Pastorello A., Gal-Yam A., Fox D.~B., Leonard D.~C., et al., 2008, MNRAS, 391, L5

\bibitem[\protect\citeauthoryear{Crowther}{2007}]{Crowther2007} Crowther P.~A., 2007, ARA\&A, 45, 177

\bibitem[\protect\citeauthoryear{Crowther}{2013}]{Crowther2013} Crowther P.~A., 2013, MNRAS, 428, 1927

\bibitem[\protect\citeauthoryear{della Valle et al.}{1990}]{sn1990w.ref1} della Valle M., Pasquini L., Phillips M., McCarthy P., 1990, IAUC, 5079

\bibitem[\protect\citeauthoryear{Dessart et al.}{2012}]{Dessart2012} Dessart L., Hillier D.~J., Li C., Woosley S., 2012, MNRAS, 424, 2139

\bibitem[\protect\citeauthoryear{Dessart, Hillier, \& Kuncarayakti}{2022}]{Dessart2022} Dessart L., Hillier D. J., Kuncarayakti H., 2022, A\&A, 658, A130

\bibitem[\protect\citeauthoryear{Dolphin}{2000}]{dolphot.ref} Dolphin A.~E., 2000, PASP, 112, 1383

\bibitem[\protect\citeauthoryear{Drout et al.}{2016}]{sn2013ge.ref} Drout M.~R., Milisavljevic D., Parrent J., Margutti R., Kamble A., Soderberg A.~M., Challis P., et al., 2016, ApJ, 821, 57

\bibitem[\protect\citeauthoryear{Efremov}{1995}]{Efremov1995} Efremov Y.~N., 1995, AJ, 110, 2757

\bibitem[\protect\citeauthoryear{Eldridge et al.}{2015}]{Eldridge2015} Eldridge J.~J., Fraser M., Maund J.~R., Smartt S.~J., 2015, MNRAS, 446, 2689

\bibitem[\protect\citeauthoryear{Fang et al.}{2019}]{Fang2019} Fang Q., Maeda K., Kuncarayakti H., Sun F., Gal-Yam A., 2019, NatAs, 3, 434

\bibitem[\protect\citeauthoryear{Fang et al.}{2022}]{Fang2022} Fang Q., Maeda K., Kuncarayakti H., Tanaka M., Kawabata K.~S., Hattori T., Aoki K., et al., 2022, ApJ, 928, 151

\bibitem[\protect\citeauthoryear{Filippenko \& McCarthy}{1990}]{sn1990w.ref2} Filippenko A.~V., McCarthy P., 1990, IAUC, 5090

\bibitem[\protect\citeauthoryear{Filippenko \& Korth}{1991}]{sn1991n.ref} Filippenko A.~V., Korth S., 1991, IAUC, 5234

\bibitem[\protect\citeauthoryear{Filippenko \& Chornock}{2000}]{sn2000ds.ref} Filippenko A.~V., Chornock R., 2000, IAUC, 7511

\bibitem[\protect\citeauthoryear{Filippenko, Chornock, \& Modjaz}{2000}]{sn2000ew.ref} Filippenko A.~V., Chornock R., Modjaz M., 2000, IAUC, 7547

\bibitem[\protect\citeauthoryear{Filippenko \& Chornock}{2001}]{sn2001ci.ref} Filippenko A.~V., Chornock R., 2001, IAUC, 7638

\bibitem[\protect\citeauthoryear{Filippenko \& Foley}{2005}]{sn2005ae.ref} Filippenko A.~V., Foley R.~J., 2005, IAUC, 8486

\bibitem[\protect\citeauthoryear{Fitzpatrick}{2004}]{avlaw.ref} Fitzpatrick E.~L., 2004, ASPC, 309, 33, ASPC..309

\bibitem[\protect\citeauthoryear{Folatelli et al.}{2014}]{sn2010as.ref} Folatelli G., Bersten M.~C., Kuncarayakti H., Olivares Estay F., Anderson J.~P., Holmbo S., Maeda K., et al., 2014, ApJ, 792, 7

\bibitem[\protect\citeauthoryear{Foley et al.}{2004}]{sn2004bm.ref} Foley R.~J., Wong D.~S., Ganeshalingam M., Filippenko A.~V., Chornock R., 2004, IAUC, 8339

\bibitem[\protect\citeauthoryear{Foley et al.}{2007}]{sn2006jc.ref1} Foley R.~J., Smith N., Ganeshalingam M., Li W., Chornock R., Filippenko A.~V., 2007, ApJL, 657, L105

\bibitem[\protect\citeauthoryear{Foley et al.}{2009}]{sn2008ha.ref} Foley R.~J., Chornock R., Filippenko A.~V., Ganeshalingam M., Kirshner R.~P., Li W., Cenko S.~B., et al., 2009, AJ, 138, 376

\bibitem[\protect\citeauthoryear{Foley}{2009}]{sn2009gj.ref} Foley R.~J., 2009, CBET, 1858

\bibitem[\protect\citeauthoryear{Fox et al.}{2022}]{Fox2022} Fox O.~D., Van Dyk S.~D., Williams B.~F., Drout M., Zapartas E., Smith N., Milisavljevic D., et al., 2022, ApJL, 929, L15

\bibitem[\protect\citeauthoryear{Fuller \& Ro}{2018}]{Fuller2018} Fuller J., Ro S., 2018, MNRAS, 476, 1853

\bibitem[\protect\citeauthoryear{Gagliano et al.}{2022}]{Gagliano2022} Gagliano A., Izzo L., Kilpatrick C.~D., Mockler B., Jacobson-Gal{\'a}n W.~V., Terreran G., Dimitriadis G., et al., 2022, ApJ, 924, 55

\bibitem[\protect\citeauthoryear{Galbany et al.}{2018}]{Galbany2018} Galbany L., Anderson J.~P., S{\'a}nchez S.~F., Kuncarayakti H., Pedraz S., Gonz{\'a}lez-Gait{\'a}n S., Stanishev V., et al., 2018, ApJ, 855, 107

\bibitem[\protect\citeauthoryear{Green}{2009}]{sn2009dq.ref} Green D.~W.~E., 2009, CBET, 1789

\bibitem[\protect\citeauthoryear{Green}{2009}]{sn2009em.ref1} Green D.~W.~E., 2009, CBET, 1807

\bibitem[\protect\citeauthoryear{Hamuy et al.}{2009}]{sn2003bg.ref1} Hamuy M., Deng J., Mazzali P.~A., Morrell N.~I., Phillips M.~M., Roth M., Gonzalez S., et al., 2009, ApJ, 703, 1612

\bibitem[\protect\citeauthoryear{Heger et al.}{2003}]{Heger2003} Heger A., Fryer C.~L., Woosley S.~E., Langer N., Hartmann D.~H., 2003, ApJ, 591, 288

\bibitem[\protect\citeauthoryear{Hirai et al.}{2020}]{Hirai2020} Hirai R., Sato T., Podsiadlowski P., Vigna-G{\'o}mez A., Mandel I., 2020, MNRAS, 499, 1154

\bibitem[\protect\citeauthoryear{Hosseinzadeh et al.}{2016a}]{sn2016bas.ref} Hosseinzadeh G., Arcavi I., Howell D.~A., McCully C., Valenti S., 2016a, ATel, 8814

\bibitem[\protect\citeauthoryear{Hosseinzadeh et al.}{2016b}]{sn2016cdd.ref} Hosseinzadeh G., Arcavi I., Howell D.~A., McCully C., Valenti S., 2016b, ATel, 9065

\bibitem[\protect\citeauthoryear{Itagaki et al.}{2012}]{sn2012cw.ref} Itagaki K., Noguchi T., Nakano S., Yusa T., Wang X.-F., Liu Q., Zhang J.-J., et al., 2012, CBET, 3148

\bibitem[\protect\citeauthoryear{Jeffreys}{1961}]{bayes.ref} Jeffreys H., 1961, Theory of Probability, third edition. Oxford University Press, Oxford

\bibitem[\protect\citeauthoryear{Kangas et al.}{2013}]{Kangas2013} Kangas T., Mattila S., Kankare E., Kotilainen J.~K., V{\"a}is{\"a}nen P., Greimel R., Takalo A., 2013, MNRAS, 436, 3464

\bibitem[\protect\citeauthoryear{Kankare et al.}{2014}]{sn2005at.ref1} Kankare E., Fraser M., Ryder S., Romero-Ca{\~n}izales C., Mattila S., Kotak R., Laursen P., et al., 2014, A\&A, 572, A75

\bibitem[\protect\citeauthoryear{Kim et al.}{2014}]{sn2014c.ref1} Kim M., Zheng W., Li W., Filippenko A.~V., Cenko S.~B., Arbour R., Masi G., et al., 2014, CBET, 3777

\bibitem[\protect\citeauthoryear{Kilpatrick et al.}{2017}]{Kilpatrick2017} Kilpatrick C.~D., Foley R.~J., Abramson L.~E., Pan Y.-C., Lu C.-X., Williams P., Treu T., et al., 2017, MNRAS, 465, 4650

\bibitem[\protect\citeauthoryear{Kilpatrick et al.}{2021}]{Kilpatrick2021} Kilpatrick C.~D., Drout M.~R., Auchettl K., Dimitriadis G., Foley R.~J., Jones D.~O., DeMarchi L., et al., 2021, MNRAS, 504, 2073

\bibitem[\protect\citeauthoryear{Kinugasa et al.}{2002}]{sn2002ap.ref} Kinugasa K., Kawakita H., Ayani K., Kawabata T., Yamaoka H., Deng J.~S., Mazzali P.~A., et al., 2002, ApJL, 577, L97

\bibitem[\protect\citeauthoryear{Koenig et al.}{2008}]{Koenig2008} Koenig X.~P., Allen L.~E., Gutermuth R.~A., Hora J.~L., Brunt C.~M., Muzerolle J., 2008, ApJ, 688, 1142

\bibitem[\protect\citeauthoryear{Kuncarayakti et al.}{2018}]{K2018} Kuncarayakti H., Anderson J.~P., Galbany L., Maeda K., Hamuy M., Aldering G., Arimoto N., et al., 2018, A\&A, 613, A35

\bibitem[\protect\citeauthoryear{Lane et al.}{1995}]{sn1995f.ref} Lane D.~J., Gray P., Fillipenko A.~V., Barth A.~J., 1995, IAUC, 6138

\bibitem[\protect\citeauthoryear{Leloudas et al.}{2012}]{sn2006oz.ref} Leloudas G., Chatzopoulos E., Dilday B., Gorosabel J., Vinko J., Gallazzi A., Wheeler J.~C., et al., 2012, A\&A, 541, A129

\bibitem[\protect\citeauthoryear{Lyman et al.}{2016}]{Lyman2016} Lyman J.~D., Bersier D., James P.~A., Mazzali P.~A., Eldridge J.~J., Fraser M., Pian E., 2016, MNRAS, 457, 328

\bibitem[\protect\citeauthoryear{Makarov, et al.}{2014}]{HyperLEDA.ref} Makarov D., Prugniel P., Terekhova N., Courtois H., Vauglin I., 2014, A\&A, 570, A13

\bibitem[\protect\citeauthoryear{Martins \& Palacios}{2013}]{Martins2013} Martins F., Palacios A., 2013, A\&A, 560, A16

\bibitem[\protect\citeauthoryear{Maund et al.}{2004a}]{Maund2004} Maund J.~R., Smartt S.~J., Kudritzki R.~P., Podsiadlowski P., Gilmore G.~F., 2004, Natur, 427, 129

\bibitem[\protect\citeauthoryear{Matheson et al.}{2004b}]{sn2004ao.ref} Matheson T., Challis P., Kirshner R., Berlind P., 2004, IAUC, 8304

\bibitem[\protect\citeauthoryear{Maund et al.}{2011}]{Maund2011} Maund J.~R., Fraser M., Ergon M., Pastorello A., Smartt S.~J., Sollerman J., Benetti S., et al., 2011, ApJL, 739, L37

\bibitem[\protect\citeauthoryear{Maund et al.}{2016}]{Maund2016b} Maund J.~R., Pastorello A., Mattila S., Itagaki K., Boles T., 2016, ApJ, 833, 128

\bibitem[\protect\citeauthoryear{Maund \& Ramirez-Ruiz}{2016}]{Maund2016} Maund J.~R., Ramirez-Ruiz E., 2016, MNRAS, 456, 3175

\bibitem[\protect\citeauthoryear{Maund}{2017}]{Maund2017} Maund J.~R., 2017, MNRAS, 469, 2202

\bibitem[\protect\citeauthoryear{Maund}{2018}]{Maund2018} Maund J.~R., 2018, MNRAS, 476, 2629

\bibitem[\protect\citeauthoryear{Maund}{2019}]{Maund2019} Maund J.~R., 2019, ApJ, 883, 86

\bibitem[\protect\citeauthoryear{Mazzali et al.}{2002}]{Mazzali2002} Mazzali P.~A., Deng J., Maeda K., Nomoto K., Umeda H., Hatano K., Iwamoto K., et al., 2002, ApJL, 572, L61

\bibitem[\protect\citeauthoryear{Mazzali et al.}{2004}]{Mazzali2004} Mazzali P.~A., Deng J., Maeda K., Nomoto K., Filippenko A.~V., Matheson T., 2004, ApJ, 614, 858

\bibitem[\protect\citeauthoryear{Mazzali et al.}{2004}]{sn1997dq.ref} Mazzali P.~A., Deng J., Maeda K., Nomoto K., Filippenko A.~V., Matheson T., 2004, ApJ, 614, 858

\bibitem[\protect\citeauthoryear{Mazzali et al.}{2009}]{sn2003bg.ref2} Mazzali P.~A., Deng J., Hamuy M., Nomoto K., 2009, ApJ, 703, 1624

\bibitem[\protect\citeauthoryear{Milisavljevic et al.}{2011}]{sn2011hs.ref} Milisavljevic D., Fesen R., Soderberg A., Pickering T., Kotze P., 2011, CBET, 2902

\bibitem[\protect\citeauthoryear{Milisavljevic et al.}{2015}]{sn2014c.ref2} Milisavljevic D., Margutti R., Kamble A., Patnaude D.~J., Raymond J.~C., Eldridge J.~J., Fong W., et al., 2015, ApJ, 815, 120

\bibitem[\protect\citeauthoryear{Miller et al.}{2022}]{Miller2022} Miller A.~E., Cioni M.-R.~L., de Grijs R., Sun N.-C., Bell C.~P.~M., Choudhury S., Ivanov V.~D., et al., 2022, MNRAS, 512, 1196. 

\bibitem[\protect\citeauthoryear{Modjaz et al.}{2005}]{sn2005ar.ref} Modjaz M., Kirshner R., Challis P., Matheson T., Foster J., 2005, IAUC, 8493

\bibitem[\protect\citeauthoryear{Modjaz et al.}{2020}]{Modjaz2020} Modjaz M., Bianco F.~B., Siwek M., Huang S., Perley D.~A., Fierroz D., Liu Y.-Q., et al., 2020, ApJ, 892, 153

\bibitem[\protect\citeauthoryear{Monard et al.}{2014}]{sn2014df.ref} Monard L.~A.~G., Kneip R., Brimacombe J., Sato H., Childress M., Zhou G., Scalzo R., et al., 2014, CBET, 3977

\bibitem[\protect\citeauthoryear{Morrell et al.}{2005}]{sn2005aw.ref} Morrell N., Hamuy M., Folatelli G., Roth M., 2005, IAUC, 8507

\bibitem[\protect\citeauthoryear{Nakano et al.}{2012}]{sn2012fh.ref} Nakano S., Yusa T., Yoshimoto K., Tomasella L., Turatto M., Pastorello A., Ochner P., et al., 2012, CBET, 3263

\bibitem[\protect\citeauthoryear{Navasardyan et al.}{2008}]{sn2008bo.ref} Navasardyan H., Benetti S., Harutyunyan A., Agnoletto I., Bufano F., Cappellaro E., Turatto M., 2008, CBET, 1325

\bibitem[\protect\citeauthoryear{Navasardyan \& Benetti}{2009}]{sn2009em.ref2} Navasardyan H., Benetti S., 2009, CBET, 1806

\bibitem[\protect\citeauthoryear{Pastorello et al.}{2007}]{sn2006jc.ref2} Pastorello A., Smartt S.~J., Mattila S., Eldridge J.~J., Young D., Itagaki K., Yamaoka H., et al., 2007, Natur, 447, 829

\bibitem[\protect\citeauthoryear{Podsiadlowski, Joss, \& Hsu}{1992}]{Pods1992} Podsiadlowski P., Joss P.~C., Hsu J.~J.~L., 1992, ApJ, 391, 246

\bibitem[\protect\citeauthoryear{Qiu et al.}{1999}]{sn1996cb.ref} Qiu Y., Li W., Qiao Q., Hu J., 1999, AJ, 117, 736

\bibitem[\protect\citeauthoryear{Quimby et al.}{2004}]{sn2004gk.ref} Quimby R., Gerardy C., Hoeflich P., Wheeler J.~C., Shetrone M., Riley V., McGaha J., et al., 2004, IAUC, 8446

\bibitem[\protect\citeauthoryear{Richardson et al.}{2022}]{Richardson2022} Richardson C.~J., Zanolin M., Andresen H., Szczepa{\'n}czyk M.~J., Gill K., Wongwathanarat A., 2022, PhRvD, 105, 103008


\bibitem[\protect\citeauthoryear{Ryder et al.}{2018}]{Ryder2018} Ryder S.~D., Van Dyk S.~D., Fox O.~D., Zapartas E., de Mink S.~E., Smith N., Brunsden E., et al., 2018, ApJ, 856, 83

\bibitem[\protect\citeauthoryear{Salpeter}{1955}]{imf.ref} Salpeter E.~E., 1955, ApJ, 121, 161

\bibitem[\protect\citeauthoryear{Schady et al.}{2019}]{Schady2019} Schady P., Eldridge J.~J., Anderson J., Chen T.-W., Galbany L., Kuncarayakti H., Xiao L., 2019, MNRAS, 490, 4515

\bibitem[\protect\citeauthoryear{Schlafly \& Finkbeiner}{2011}]{Schlafly2011} Schlafly E.~F., Finkbeiner D.~P., 2011, ApJ, 737, 103

\bibitem[\protect\citeauthoryear{Schmidt \& Salvo}{2005}]{sn2005at.ref2} Schmidt B., Salvo M., 2005, IAUC, 8496

\bibitem[\protect\citeauthoryear{Schneider et al.}{2015}]{Schneider2015} Schneider F.~R.~N., Izzard R.~G., Langer N.,     de Mink S.~E., 2015, ApJ, 805, 20. doi:10.1088/0004-637X/805/1/20

\bibitem[\protect\citeauthoryear{Shivvers et al.}{2017}]{sn2004c.ref} Shivvers I., Modjaz M., Zheng W., Liu Y., Filippenko A.~V., Silverman J.~M., Matheson T., et al., 2017, PASP, 129, 054201

\bibitem[\protect\citeauthoryear{Shivvers et al.}{2016}]{sn2015u.ref} Shivvers I., Zheng W.~K., Mauerhan J., Kleiser I.~K.~W., Van Dyk S.~D., Silverman J.~M., Graham M.~L., et al., 2016, MNRAS, 461, 3057

\bibitem[\protect\citeauthoryear{Shivvers et al.}{2017}]{sn2015g.ref} Shivvers I., Zheng W., Van Dyk S.~D., Mauerhan J., Filippenko A.~V., Smith N., Foley R.~J., et al., 2017, MNRAS, 471, 4381

\bibitem[\protect\citeauthoryear{Smartt}{2009}]{Smartt2009} Smartt S.~J., 2009, ARA\&A, 47, 63

\bibitem[\protect\citeauthoryear{Smith}{2014}]{Smith2014} Smith N., 2014, ARA\&A, 52, 487

\bibitem[\protect\citeauthoryear{Smith}{2017}]{Smith2017}Smith N., 2017, Handbook of Supernovae, Springer International Publishing AG, p. 403

\bibitem[\protect\citeauthoryear{Sollerman, Leibundgut, \& Spyromilio}{1998}]{sn1996n.ref} Sollerman J., Leibundgut B., Spyromilio J., 1998, A\&A, 337, 207

\bibitem[\protect\citeauthoryear{Sun et al.}{2017a}]{Sun2017a} Sun N.-C., de Grijs R., Subramanian S., Cioni M.-R.~L., Rubele S., Bekki K., Ivanov V.~D., et al., 2017, ApJ, 835, 171

\bibitem[\protect\citeauthoryear{Sun et al.}{2017b}]{Sun2017b} Sun N.-C., de Grijs R., Subramanian S., Bekki K., Bell C.~P.~M., Cioni M.-R.~L., Ivanov V.~D., et al., 2017, ApJ, 849, 149

\bibitem[\protect\citeauthoryear{Sun et al.}{2018}]{Sun2018} Sun N.-C., de Grijs R., Cioni M.-R.~L., Rubele S., Subramanian S., van Loon J.~T., Bekki K., et al., 2018, ApJ, 858, 31

\bibitem[\protect\citeauthoryear{Sun et al.}{2020}]{Sun2020a} Sun N.-C., Maund J.~R., Hirai R., Crowther P.~A., Podsiadlowski P., 2020, MNRAS, 491, 6000

\bibitem[\protect\citeauthoryear{Sun, Maund, \& Crowther}{2020}]{Sun2020b} Sun N.-C., Maund J.~R., Crowther P.~A., 2020, MNRAS, 497, 5118

\bibitem[\protect\citeauthoryear{Sun et al.}{2021}]{Sun2021} Sun N.-C., Maund J.~R., Crowther P.~A., Fang X., Zapartas E., 2021, MNRAS, 504, 2253

\bibitem[\protect\citeauthoryear{Sun et al.}{2022}]{Sun2022a} Sun N.-C., Maund J.~R., Crowther P.~A., Hirai R., Kashapov A., Liu J.-F., Liu L.-D., et al., 2022a, MNRAS, 510, 3701


\bibitem[\protect\citeauthoryear{Taddia et al.}{2019}]{Taddia2019} Taddia F., Sollerman J., Fremling C., Barbarino C., Karamehmetoglu E., Arcavi I., Cenko S.~B., et al., 2019, A\&A, 621, A71

\bibitem[\protect\citeauthoryear{Tartaglia et al.}{2017}]{Tartaglia2017} Tartaglia L., Fraser M., Sand D.~J., Valenti S., Smartt S.~J., McCully C., Anderson J.~P., et al., 2017, ApJL, 836, L12

\bibitem[\protect\citeauthoryear{Taubenberger et al.}{2005}]{sn2005v.ref2} Taubenberger S., Pastorello A., Benetti S., Aceituno J., 2005, IAUC, 8474


\bibitem[\protect\citeauthoryear{Tully, et al.}{2013}]{CosmicFlow2.ref} Tully R.~B., et al., 2013, AJ, 146, 86

\bibitem[\protect\citeauthoryear{Tully, Courtois \& Sorce}{2016}]{CosmicFlow3.ref} Tully R.~B., Courtois H.~M., Sorce J.~G., 2016, AJ, 152, 50

\bibitem[\protect\citeauthoryear{Van Dyk et al.}{2018}]{VanDyk2018} Van Dyk S.~D., Zheng W., Brink T.~G., Filippenko A.~V., Milisavljevic D., Andrews J.~E., Smith N., et al., 2018, ApJ, 860, 90

\bibitem[\protect\citeauthoryear{Van Dyk et al.}{2014}]{sn2013df.ref} Van Dyk S.~D., Zheng W., Fox O.~D., Cenko S.~B., Clubb K.~I., Filippenko A.~V., Foley R.~J., et al., 2014, AJ, 147, 37

\bibitem[\protect\citeauthoryear{Vink}{2017}]{Vink2017} Vink J.~S., 2017, A\&A, 607, L8

\bibitem[\protect\citeauthoryear{Wiggins et al.}{2015}]{sn2015q.ref} Wiggins P., Masi G., Brimacombe J., Belligoli R., Castellani F., Marangoni C., Leonard D.~C., et al., 2015, CBET, 4128

\bibitem[\protect\citeauthoryear{Williams et al.}{2019}]{Williams2019} Williams B.~F., Hillis T.~J., Blair W.~P., Long K.~S., Murphy J.~W., Dolphin A., Khan R., et al., 2019, ApJ, 881, 54

\bibitem[\protect\citeauthoryear{Wu \& Fuller}{2022}]{Wu2022} Wu S.~C., Fuller J., 2022, ApJ, 930, 119

\bibitem[\protect\citeauthoryear{Xiang et al.}{2019}]{Xiang2019} Xiang D., Wang X., Mo J., Wang L., Smartt S., Fraser M., Ehgamberdiev S.~A., et al., 2019, ApJ, 871, 176

\bibitem[\protect\citeauthoryear{Xiao, Stanway, \& Eldridge}{2018}]{Xiao2018} Xiao L., Stanway E.~R., Eldridge J.~J., 2018, MNRAS, 477, 904

\bibitem[\protect\citeauthoryear{Xiao et al.}{2019}]{Xiao2019} Xiao L., Galbany L., Eldridge J.~J., Stanway E.~R., 2019, MNRAS, 482, 384

\bibitem[\protect\citeauthoryear{Yoon}{2017}]{Yoon2017b} Yoon S.-C., 2017, MNRAS, 470, 3970

\bibitem[\protect\citeauthoryear{Yoon, Dessart, \& Clocchiatti}{2017}]{Yoon2017} Yoon S.-C., Dessart L., Clocchiatti A., 2017, ApJ, 840, 10

\bibitem[\protect\citeauthoryear{Zapartas et al.}{2017}]{Zapartas2017} Zapartas E., de Mink S.~E., Van Dyk S.~D., Fox O.~D., Smith N., Bostroem K.~A., de Koter A., et al., 2017, ApJ, 842, 125

\end{thebibliography}
\end{document}